\documentclass[%
reprint,
superscriptaddress,
 amsmath,amssymb,
 aps,
 pra,
]{revtex4-2}

\usepackage{dcolumn}% Align table columns on decimal point
\usepackage{bm}% bold math
\usepackage{graphicx} %23/12/22 additional
\begin{document}
 
\preprint{APS/123-QED}

\title{EP restoration and fast-light edge states in photonic crystal waveguide\\
with glide and time reversal symmetry
}% Force line breaks with \\

\author{Takahiro Uemura}
\affiliation{Department of Physics, Tokyo Institute of Technology, 2-12-1 Ookayama, Meguro-ku, Tokyo 152-8551, Japan}
\affiliation{NTT Basic Research Laboratories, NTT Corporation, 3-1 Morinosato-Wakamiya, Atsugi-shi, Kanagawa 243-0198, Japan}

\author{Taiki Yoda}
\affiliation{Department of Physics, Tokyo Institute of Technology, 2-12-1 Ookayama, Meguro-ku, Tokyo 152-8551, Japan}

\author{Yuto Moritake}
\affiliation{Department of Physics, Tokyo Institute of Technology, 2-12-1 Ookayama, Meguro-ku, Tokyo 152-8551, Japan}

\author{Shutaro Otsuka}
\affiliation{Department of Physics, Tokyo Institute of Technology, 2-12-1 Ookayama, Meguro-ku, Tokyo 152-8551, Japan}
\affiliation{NTT Basic Research Laboratories, NTT Corporation, 3-1 Morinosato-Wakamiya, Atsugi-shi, Kanagawa 243-0198, Japan}

\author{Kenta Takata}
\affiliation{NTT Basic Research Laboratories, NTT Corporation, 3-1 Morinosato-Wakamiya, Atsugi-shi, Kanagawa 243-0198, Japan}
\affiliation{Nanophotonics Center, NTT Corporation, 3-1, Morinosato-Wakamiya, Atsugi-shi, Kanagawa 243-0198, Japan}

\author{Masaya Notomi}
\affiliation{Department of Physics, Tokyo Institute of Technology, 2-12-1 Ookayama, Meguro-ku, Tokyo 152-8551, Japan}
\affiliation{NTT Basic Research Laboratories, NTT Corporation, 3-1 Morinosato-Wakamiya, Atsugi-shi, Kanagawa 243-0198, Japan}
\affiliation{Nanophotonics Center, NTT Corporation, 3-1, Morinosato-Wakamiya, Atsugi-shi, Kanagawa 243-0198, Japan}
\email{notomi@phys.titech.ac.jp}


\begin{abstract}
Exceptional points (EPs) in the propagation states give rise to 
the emergence of intriguing properties with
the divergence of the group velocity.
However, there have been no experimental reports 
due to the necessity of maintaining high levels of fabrication precision 
and the requisite high group velocity contrast. 
In our study, we propose a design of photonic crystal waveguide with glide and time reversal symmetry,
and derive an effective Hamiltonian for edge states to realize fast-light edge states. 
We adopt a systematic method to generate EPs in edge states 
by introducing non-Hermitian perturbations to Dirac points guaranteed by glide symmetry,
which ensures that EP modes are free from out-of-plane radiation losses.
Then, our study reveals the conditions for the exact EP restoration 
and provides an analytical solution to offset the EP smoothing due to symmetry breaking,
which drastically reduces the group velocity contrast.
A good symmetry property of the photonic crystal waveguide allows us to derive the effective Hamiltonian as a simple form, and the EPs can be restored by adjusting the real part of the permittivity.
Furthermore, we design a feasible photonic crystal slab waveguide 
incorporating graphene as the absorbing material, 
and numerically demonstrate a group velocity reaching $v_g = 3.3c$ near the EP, 
which is up to 25 times that of the original structure.
Thanks to the short periodicity of photonic crystals, 
it's possible to reach the speed of light in vacuum 
with group velocity contrasts on the order of one digit. 
Our study paves an innovative way to manipulate the group velocity of light.
\end{abstract}

\maketitle

%\tableofcontents

\section{Introduction}

Non-Hermitian optical systems have gained interest as a new approach for controlling light. 
PT-symmetric systems, which exhibit symmetry in spatial parity and time inversion, 
have received considerable attention \cite{PhysRevLett.80.5243, Bender_2007}. 
Above all, PT-symmetric optical systems are significant for implementing complex potentials 
through the use of optical gain from lasers \cite{PhysRevLett.120.113901} or current injection \cite{Takata:21}, 
coupled with loss from radiation or material absorption 
\cite{Feng2017, Regensburger2012, Ozdemir2019}. 
PT-symmetric systems have been used to realize various optical phenomena,
including non-reciprocal transmission \cite{Peng2014}, 
single-mode lasing \cite{doi:10.1126/science.1258479, doi:10.1126/science.1258480}
and loss-induced transparency \cite{PhysRevLett.103.093902}.

In the context of periodic potential structures, 
it has been noted that the slope of the band diverges 
in the vicinity of the exceptional point (EP) \cite{PhysRevA.84.021806}.
The control of light propagation stands as a fundamental concern in photonics, 
and the manipulation of the group velocity of light is a central technology 
for the advancement of optical devices. 
Recent advancements in semiconductor microfabrication technology have enabled 
the realization of on-chip slow light in nanophotonics \cite{Baba2008},
including resonator-type coupled resonator optical waveguides (CROW) \cite{Yariv:99, Olivier:01, Notomi2008}
and photonic crystal waveguide \cite{Vlasov2005, Yoshimi:20, Yoshimi:21}.
Conversely, light propagating with a group velocity faster than the speed of light in vacuum, 
known as fast light, has been observed in media 
with anomalous dispersion near an absorption line \cite{Brillouin:1960tos},
atomic systems \cite{Wang2000, doi:10.1126/science.1084429}, 
as well as in artificial media like metamaterials \cite{PhysRevE.63.046604, PhysRevLett.109.223903}. 
However, achieving fast light in conventional transparent media
for on-chip waveguides remains a challenge.
The concept of PT-symmetric optical waveguide system harbors the potential 
to bring about a breakthrough in achieving fast light in on-chip waveguides.

While such interesting properties in PT-symmetric optical waveguide system have been predicted,
the realization of fast light in PT-symmetric systems remain a challenge.
The anomalous group velocity near EPs, referred to as fast light, has been discussed in Ref. \cite{PhysRevA.88.052102, PhysRevA.90.053819}, and Ref. \cite{PhysRevApplied.7.054023} proposes and theoretically investigates 
fast light in PT-symmetric systems 
and implements it numerically in CROW structures using photonic crystal cavity arrays.
However, the low group velocity and strong light confinement make it difficult to apply CROW to fast-light propagation states.
The initial group velocities of the mode in CROW are inherently slow
(e.g. $v_g=0.02c$ with the periods of cavities 2.1 $\mu$m \cite{PhysRevApplied.7.054023}).
That is, to achieve group velocities exceeding the speed of light in a vacuum 
requires about an order of $100 v_g$, 
which requires extremely precise control of spatial structures as well as gains and losses.
Moreover, the strong light confinement in the cavity makes it increasingly difficult 
to achieve high fabrication precision. 
The issues related to group velocity and fabrication precision can be resolved 
by weakening the light confinement to increase the coupling rate between resonators, 
even when the periods of cavities is large. 
However, this approach is contrary to the typical CROW configuration. 
When we pursue CROWs with wide periods and weak light confinement, 
the CROW structure can be regarded as a photonic crystal with a large period. 
In this case, we would be using very high-order bands as a photonic crystal, 
which is not advantageous compared to conventional PT-symmetric photonic crystal waveguides.
In addition to these, it should be noted that the waveguide modes of CROW are positioned above the light line, and their influence depends on the Q factor. That is, it is inherently difficult to eliminate the effects of out-of-plane radiation losses from the waveguide modes of CROW.

Therefore, our research motivation lies in realizing superluminal states 
with a group velocity contrast of one digit from the initial edge states
by exploiting the PT symmetry in photonic crystal waveguide systems,
which have much shorter periods of the order of sub-micrometers.
The photonic crystal waveguides offer additional advantages over CROW 
in terms of scalability and device integration, 
making them favorable for applications in optical devices. 
However, adopting photonic crystal band systems presents challenges not encountered in CROW, where  the smoothing of EPs due to finite Q can be canceled using a tight-binding-based formalism \cite{PhysRevA.105.013523}.
When creating EPs in photonic crystal systems, unlike in tight-binding systems, it is not known how to restore the EPs when they are smoothed by symmetry breaking. 
Therefore, the establishment of a method to restore EPs in photonic crystal systems is a significant step towards the realization of superluminal propagation states.
Since the use of photonic crystal bulk modes as waveguide modes is not realistic, 
the focus of our study is on edge states in photonic crystal waveguides.
Several methods for realizing photonic crystal waveguides have been proposed, 
one that utilizes a degenerate point created by imaginary perturbations and unit cell enlargement \cite{PhysRevB.100.115412, Mock:20};
another that employs an accidental degenerate point \cite{Fang_2019, Mock:20}; 
and a third method that uses symmetry-protected degenerate points \cite{Mock:20}. 
The first and second methods make it challenging to position the EP below the light line. 
In contrast, the third method allows for the EP to be positioned below the light line, 
which assures that the propagation modes near the EP are free from out-of-plane radiation loss. 
Additionally, the third method's allows for an analytical solution of the effective Hamiltonian
thanks to its good symmetry properties, 
making it easier to develop methods to restore the EP 
when the symmetry is broken,
which has not been noted in Ref. \cite{Mock:20} and is a new result we present in this paper.

In this paper, we propose a photonic crystal waveguide 
with the symmetry under simultaneous glide and time-reversal operations. 
Our scheme starts with the generation of EPs using a systematic approach 
with PT-symmetric perturbations at Dirac points, 
which appear due to the spatial glide symmetry \cite{PhysRevB.106.064304}, 
the EPs appear 
at the boundary of the Brillouin zone, 
providing two key advantages, 
both of which address issues inherent in traditional PT-symmetric photonic crystal waveguides \cite{PhysRevB.100.115412, Fang_2019, Mock:20}.
The first advantage is that 
the EPs can be positioned below the light line, 
ensuring that the propagation modes near the EP are free from out-of-plane radiation loss. 
This is in contrast to most structures proposed in previous studies, such as \cite{Mock:20}, 
where the smoothing of EP could not be avoided due to the radiation loss,
which drastically reduces the group velocity contrast.
The second advantage is that the symmetry at the boundary of the Brillouin zone allows 
the $\bm{k}\cdot \bm{p}$ Hamiltonian around the EPs to be represented in a simplified form, 
making it easier to model the response of edge states to non-Hermitian perturbations.
EPs in our waveguide are sensitive to any deformation of glide and time reversal symmetry, which presents a challenge to achieving high group velocity contrast.
Therefore, we derive the precise EP conditions for complex permittivity perturbations, 
which are essential for realizing superluminal propagation states.
In addition to the above advantages, 
it is remarkable that we adopt the valley photonic crystal waveguide as the basic structure, 
which exhibits topological properties and has potential 
for future applications in optical circuits with bends \cite{Plotnik2014}.
We also mention that our proposed structure is characterized by its ease of fabrication. 
Our design can be implemented by selectively loading graphene 
along the glide plane of a photonic crystal slab waveguide, 
which is simpler than the embedded heterostructures that require precise gain and loss adjustments.
As a whole, our study is a step 
towards experimental observations of superluminal propagation states 
in on-chip non-Hermitian nano-optical systems.

\section{Design of non-Hermitian photonic crystal waveguides with glide and time reversal symmetry}

First, we present the design of our photonic crystal heterostructure 
based on two-dimensional (2D) simulation, 
where the system extends infinitely along the $z$-axis.
All numerical calculations in this paper were performed using finite element method (FEM),
COMSOL Multiphysics.

%%%%% Figure 1 %%%%%
\begin{figure}%
    \includegraphics[width=8.4cm]{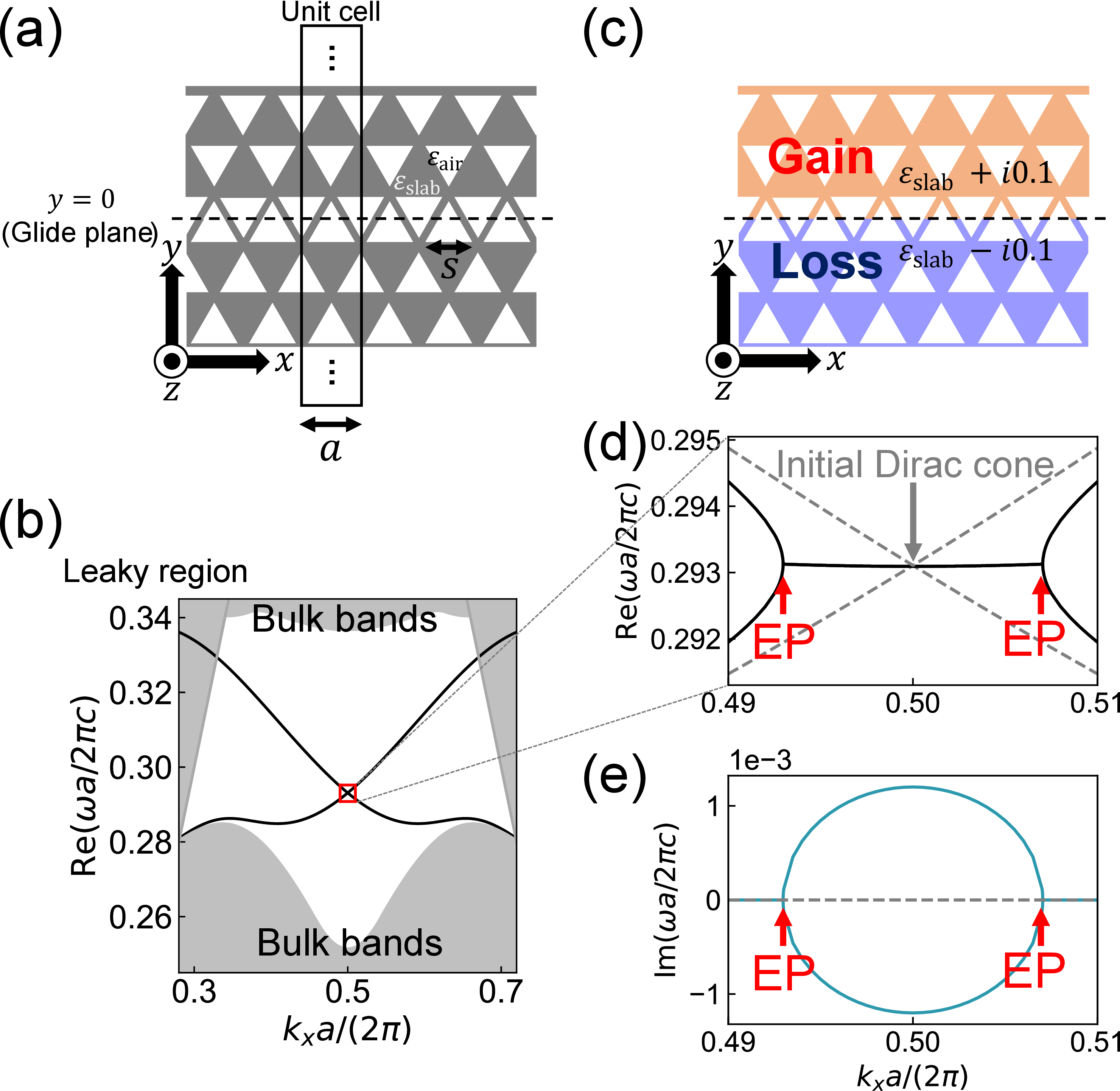}% Here is how to import EPS art
    \caption{\label{fig_2Dwg}
        (a) Interface geometry of the photonic crystal hetero-structure
        with the triangular lattice and triangular holes.
        This structure has glide symmetry along $y=0$ plane as shown by the dashed line.
        (b) The band dispersion curves in the $k_x$ direction of the TE mode in geometry (a).
        (c) Interface geometry of the glide symmetric photonic crystal hetero-structure
        with balanced gain $(y>0)$ and loss $(y<0)$.
        This structure is invariant under both a glide operation with the $xz$-plane
        and a simultaneous time-reversal operation.
        (d), (e) The real and imaginary part of band dispersion curves along the $k_x$ direction
         near $k_x a = \pi$.
    }
\end{figure}

We start from the design of a glide symmetric photonic crystal heterostructure 
with the triangular lattice and triangular holes 
as shown in Fig. \ref{fig_2Dwg}(a). 
This is also known as the bearded-edge of the photonic crystal \cite{yoda2019air}.
We considered also other types of lattices and hole shapes (See Appendix. \ref{sec:appendix_design_heterostructure}), 
but we adoped the triangular lattice and triangular holes 
because we found that Dirac points can be easily located inside the
band gap and below the light line.
The difference between triangular and circular holes lies 
in the presence or absence of valley-polarized features near the K(K') point in the bulk band, 
and there is no essential difference for constraction of EPs in edge states.
We set the lattice constant as $a = 470$ nm, the triangular hole side as $s = 0.8a$.
The air and slab permittivity are set to $\varepsilon_\mathrm{air} = 1$ 
and $\varepsilon_\mathrm{slab} = 6.76$, respectively.
The value of $\varepsilon_\mathrm{slab}$ corresponds to an effective refractive index of typical semiconductor materials such as silicon and compound semiconductor materials. 
The dashed line in Fig. \ref{fig_2Dwg}(a) represents the glide symmetric plane,
and we define it as $y=0$ and establish our coordinate axis accordingly.
The heterostructure is invariant under a glide operation with the $xz$-plane,
which is a combination of a reflection with the $xz$-plane and a translation along the $x$-axis.
The glide operator is given by
\begin{equation}
    \hat{G} = \left\{ m_y \middle| \frac{1}{2} \hat{\mathrm{x}} \right\}
    \label{eq:glide_operator}
\end{equation}
where $m_y$ and $\hat{\mathrm{x}}$ represent the mirror reflection with respect to the
$xz$-plane and a unit vector along the $x$ axis, respectively.
Let $\varepsilon(\bm{r}),\ \bm{r}={[x,y,z]}^\top$ be the permittivity 
of initial photonic crystal, 
and $\varepsilon(\bm{r})$ is invariant with respect to the glide operation:
\begin{equation}
    \hat{G}\varepsilon(\bm{r}) 
    = \varepsilon \left(x + \frac{a}{2}, -y, z \right) 
    = \varepsilon(\bm{r}).
    \label{eq:G_to_permittivity}
\end{equation}
The band dispersion curves of TE mode along the $k_x$-axis are shown in Fig. \ref{fig_2Dwg}(b). 
The region shaded in gray represents the area outside the bulk band and beyond the light line.
The existence of Dirac point at $k_x a = \pi$ in the band dispersion is guaranteed 
by the spatial glide symmetry of the photonic crystal \cite{Söllner2015, Mahmoodian:17, Yoshimi:20}

The key point of our design is 
to introduce a gain on the upper side and a loss of the same magnitude on the lower side, 
bordering the glide plane.
Figure \ref{fig_2Dwg}(c) shows the interface geometry of the photonic crystal heterostructure
with balanced gain $(y>0)$ and loss $(y<0)$.
The perturbation of imaginary part of permittivity are set to
$\Delta \varepsilon_\mathrm{i}(\bm{r}) = 0.1 \times 1_{\bm{r}\in \mathrm{slab} \ \cap\ y > 0}$ (gain) and 
$\Delta \varepsilon_\mathrm{i}(\bm{r}) = 0.1 \times 1_{\bm{r}\in \mathrm{slab} \ \cap\ y < 0}$ (loss).
Let $\hat{T}$ be the time-reversal operator, 
then the perturbed permittivity $\varepsilon' (\bm{r})=\varepsilon (\bm{r}) + \Delta \varepsilon (\bm{r})$ 
has the symmetry with respect to the simultaneous operation of $\hat{G}$ and $\hat{T}$:
\begin{align}
    (\hat{G}\hat{T})\varepsilon'(\bm{r}) 
    = (\hat{T}\hat{G})\varepsilon'(\bm{r}) 
    = \varepsilon'(\bm{r})
    \label{eq:sym_to_permittivity}
\end{align}
%
%where ${\left( \varepsilon(\bm{r})'\right)}^*$ is the complex conjugate of $\varepsilon(\bm{r})'$.
Figures \ref{fig_2Dwg}(d) and (e) show the real and imaginary part of 
the band dispersion near $k_x a = \pi$, respectively.
The Dirac point splits into two EPs when gain and loss are balanced, 
thank to the glide and time reversal (GT) symmetry. %%% definition of the gt symmetry
Note that the glide operation on our system corresponds to the parity inversion operation in the context of PT-symmetric optical systems.
The edge states undergo a phase transition between PT-symmetric and PT-broken phases, 
just as in the general PT-symmetric optical systems.
See Appendix. \ref{sec:appendix_Hz_distribution} for the details 
with the $H_z$ distribution of the edge state.
Our approach is advantageous 
because it allows for the systematic generation of EPs 
in the band dispersion of waveguide edge states. 
In addition, the initial Dirac point is located in the band gap, 
which allows us to utilize for waveguide applications. 
%Furthermore, the EPs remain unsmoothed as long as the material gain and loss are balanced,
%since the edge states near EPs do not exhibit out-of-plane radiation loss.

\section{Reconstruction of exact EPs based on perturbation theory}

%%%%% Figure 2 %%%%%
\begin{figure}
  \includegraphics[width=8.5cm]{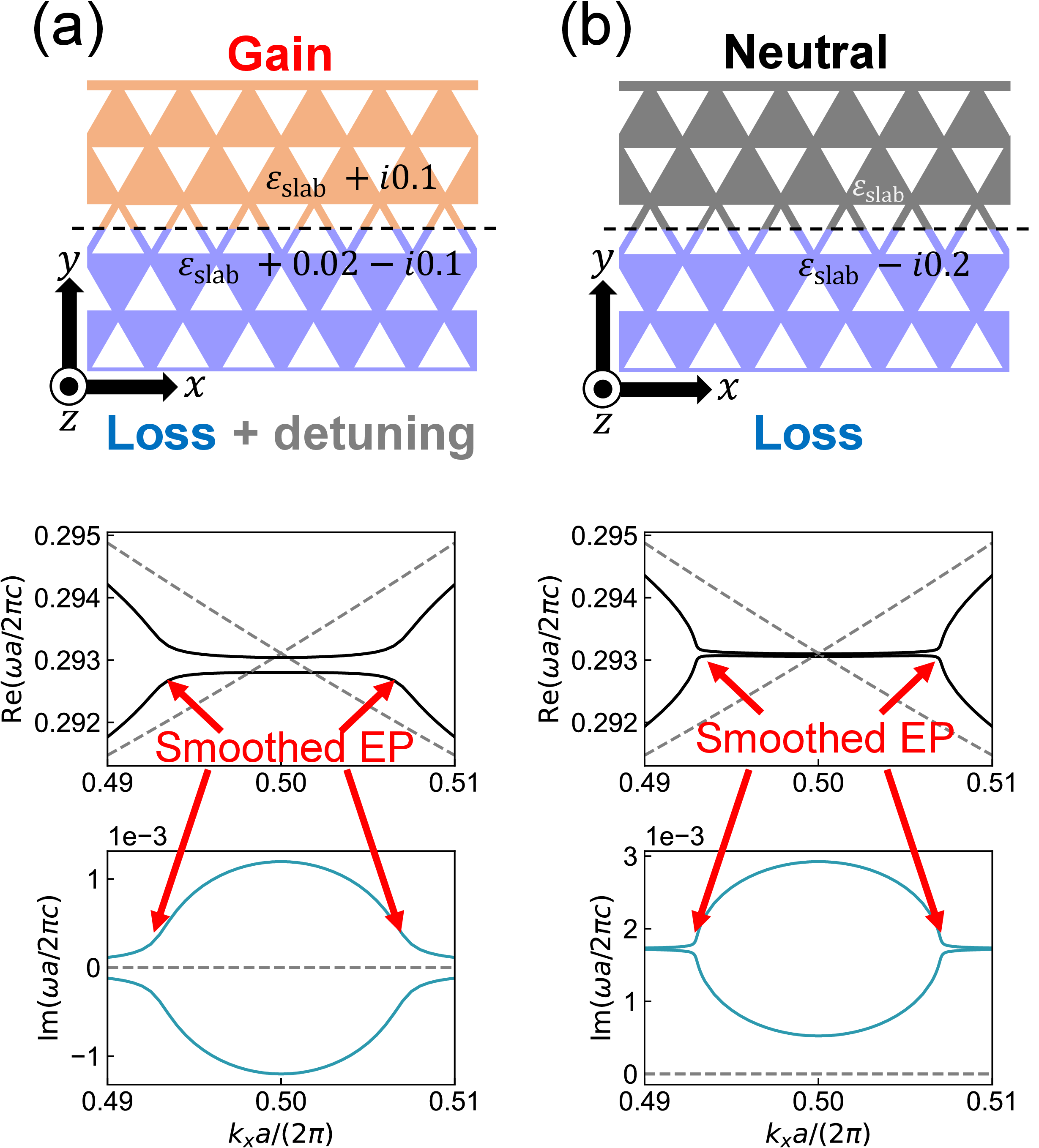}
  \caption{\label{fig_2Dlossy}
      (a) Interface geometry and corresponding band dispersion curves along the $k_x$ direction
      of the photonic crystal hetero-structure
      with symmetric gain $(y>0)$ and loss $(y<0)$, 
      but asymmetrical detuning of real part of permittivity. 
      (b) Interface geometry and corresponding band dispersion curves 
      with loss $(y<0)$.
  }
\end{figure}
In the previous section we discussed the situation with glide and time reversal symmetry.
We need to mention that the EP is sensitive to symmetry breaking; 
any detuning that breaks glide and time-reversal symmetry lifts the EP 
and smoothens the band dispersion near the EP. 
For example, as shown in Fig. \ref{fig_2Dlossy}(a), 
when a detuning of the real part of the permittivity, 
$\mathrm{Re}(\Delta \varepsilon) = 0.02$, is applied to the region of the dielectric where $y < 0$, 
the EP lifts and the band dispersion curves near the initial EP are smoothed. 
In addition, as shown in Fig. \ref{fig_2Dlossy} (b), 
introducing only loss in the dielectric region where $y < 0$ also breaks 
the glide and time-reversal symmetry due to the imbalance of gain and loss.
The ``passive'' systems with only loss are advantageous 
for the simple fabrication processes \cite{PhysRevLett.117.110802},
but it is known that the EPs are smoothed 
due to the the imaginary coupling of each eigenstate \cite{PhysRevA.105.013523}. 
In any case, the smoothing of EPs drastically reduces the group velocity contrast,
which is a critical issue for realizing superluminal propagation states.
Therefore, it is essential to develop methods 
to restore the EP by counteracting the effects of symmetry breaking.
While Ref. \cite{PhysRevA.105.013523} provides theoretical analysis and methods 
to offset the smoothing of EPs in CROW systems, 
the situation in the photonic crystal waveguide systems has not been investigated so far.

Here, we derive an effective Hamiltonian for using the $\bm{k} \cdot \bm{p}$ perturbation method to model 
the bands of edge states near the original Dirac point, i.e., $k_x a = \pi$.
Furthermore, based on the $\bm{k} \cdot \bm{p}$ Hamiltonian, 
we derive the precise EP conditions for complex permittivity perturbations,
and show that EPs can always be restored even in passive photonic crystal waveguides 
by adjusting of the real part of the permittivity.

\subsection{Effective Hamiltonian of edge states near the Dirac point}

This part describes the derivation of the effective Hamiltonian
which models the band dispersion of edge states in the photonic crystal waveguide 
near the Dirac point.
We consider the two types of perturbation,
one is the detuning of the wave vector from the Dirac point $\bm{\delta k} = [k_x - k_0, 0, 0]^\top$,
and the other denotes the permittivity perturbation $\Delta \varepsilon(\bm{r})$.
For the time being, we consider the case the permittivity perturbation is pure imaginary
$\Delta \varepsilon(\bm{r}) \in i \mathbf{R}$.
We define the two unperturbed bases 
$\bm{\varphi}_{\alpha k_0}^{(0)}(\bm{r})$ and $\bm{\psi}_{\alpha k_0}^{(0)}(\bm{r})$
as the normalized electric and magnetic fields at $k_x = \pi/a$
with $\Delta\varepsilon(\bm{r}) = 0$.
Here, $\alpha$ denotes the label of the basis,
and $C_\alpha$ denotes the expansion coefficients of the basis.
The two bases are degenerated at the Dirac point
due to the glide symmetry of the unperturbed photonic crystal.

The $\bm{k} \cdot \bm{p}$ Hamiltonian for the edge states near the Dirac point is given 
by the following $2\times 2$ matrix form (See Appendix. \ref{sec:appendix_kp_hamiltonian} for details):
\begin{align}
    \label{eq:kp_Hamiltonian_2x2}
    &\left[
    \begin{matrix}
      {\left( \frac{\omega_\alpha^{(0)}}{c} \right)}^2 + P'_{\alpha \alpha} + Q'_{\alpha \alpha} & P'_{\alpha \beta} + Q'_{\alpha \beta}  \\
      P'_{\beta \alpha} + Q'_{\beta \alpha}  & {\left( \frac{\omega_\beta^{(0)}}{c} \right)}^2 + P'_{\beta \beta} + Q'_{\beta \beta} 
    \end{matrix}
    \right]
    \left[
    \begin{matrix}
      C_\alpha \\
      C_\beta
    \end{matrix}
    \right]\nonumber\\
    &\quad \quad =
    {\left( \frac{\omega_{nk}}{c} \right)}^2
    \left[
    \begin{matrix}
      1 + G_{\alpha \alpha} & G_{\alpha \beta} \\
      G_{\beta \alpha} & 1 + G_{\beta \beta}
    \end{matrix}
    \right]
    \left[
    \begin{matrix}
      C_\alpha \\
      C_\beta
    \end{matrix}
    \right]
\end{align}
Here, $c$ is the speed of light in vacuum,
$\omega_{nk}$ represents the eigenfrequency of the $n$-th mode, 
$\alpha$ and $\beta$ denote the label of the eigenmodes,
$\omega_\alpha^{(0)}, \omega_\beta^{(0)}$ denote the unperturbed frequencies, 
and $\omega_\alpha^{(0)} = \omega_\beta^{(0)}$ holds regardless of the choice of the bases
due to the degeneracy by the glide symmetry.
$P'_{\alpha \beta}$, $G_{\alpha \beta}$ and $Q'_{\alpha \beta}$ are defined as follows:
\begin{align}
  P_{\alpha \beta}'
  &= \int_{\mathrm{u.c.}} 
  \left(
    \omega_\beta^{(0)} \bm{\psi}_{\alpha k_0}^{(0)*}(\bm{r}) \times \bm{\varphi}_{\beta k_0}^{(0)}(\bm{r})
    \right. 
    \nonumber\\
    &\qquad \qquad
    \left.
    - \omega_\alpha^{(0)} \bm{\varphi}_{\alpha k_0}^{(0)*}(\bm{r}) \times \bm{\psi}_{\beta k_0}^{(0)}(\bm{r})    
  \right) 
  \cdot \bm{\delta k} \,
  \mathrm{d}^3r
  \\
  \label{eq:def_matG}
  G_{\alpha \beta} &:=
  \int_{\mathrm{u.c.}} 
  \mathrm{d}^3r
  \Delta \varepsilon (\bm{r})
  \bm{\psi}_{\alpha k_0}^{(0)*}(\bm{r}) \cdot \bm{\psi}_{\beta k_0}^{(0)}(\bm{r})
  \\
  Q'_{\alpha \beta} &= -\int_{\mathrm{u.c.}} 
  \bm{\psi}_{\alpha k_0}^{(0)*}(\bm{r}) \cdot \left( \bm{\delta k} \times \bm{\delta k} \times \bm{\psi}_{\beta k_0}^{(0)}(\bm{r}) \right)
  \mathrm{d}^3r 
\end{align}
The $\bm{k} \cdot \bm{p}$ perturbation generally describes 
the response of frequencies and eigenstates to the perturbation of the wave vector $\bm{\delta k}$.
However, the response of the band to the perturbation 
$\Delta\varepsilon(\bm{r})$ is also naturally incorporated in the form of $G_{\alpha \beta}$.
Considering that $\bm{\delta k}$ only has a $k_x$ component, 
$P'_{\alpha \beta}$ and $Q'_{\alpha \beta}$ can be expressed using scalars $P_{\alpha \beta}$ and $Q_{\alpha \beta}$, 
which are independent of the wavenumber $\delta k_x$, as follows:
\begin{align}
  P_{\alpha \beta}' = P_{\alpha \beta} \delta k_x, \quad
  Q'_{\alpha \beta} = Q_{\alpha \beta} {(\delta k_x)}^2
\end{align}
The $P_{\alpha \beta}$ and $Q_{\alpha \beta}$ are given by
\begin{align}
  \label{eq:def_matP}
  P_{\alpha \beta}
  &= \int_{\mathrm{u.c.}} 
  \left(
    \omega_\beta^{(0)} \bm{\psi}_{\alpha k_0}^{(0)*}(\bm{r}) \times \bm{\varphi}_{\beta k_0}^{(0)}(\bm{r})
    \right. 
    \nonumber\\
    &\qquad \qquad
    \left.
    - \omega_\alpha^{(0)} \bm{\varphi}_{\alpha k_0}^{(0)*}(\bm{r}) \times \bm{\psi}_{\beta k_0}^{(0)}(\bm{r})    
  \right)_x \,
  \mathrm{d}^3r
  \\
  \label{eq:def_matQ}
  Q_{\alpha \beta} &= 
  \int_{\mathrm{u.c.}} 
  \left(
  {\left( \psi_{\alpha k_0}^{(0)*}(\bm{r}) \right)}_y
  {\left( \psi_{\beta k_0}^{(0)}(\bm{r}) \right)}_y
  \right. \nonumber\\
  &\qquad \qquad \qquad \left. + {\left( \psi_{\alpha k_0}^{(0)*}(\bm{r}) \right)}_z
  {\left( \psi_{\beta k_0}^{(0)}(\bm{r}) \right)}_z
  \right)
  \mathrm{d}^3r
\end{align}
While $G_{\alpha \beta}$ represents the effect of permittivity perturbation on the bands,
$P_{\alpha \beta}$ and $Q_{\alpha \beta}$ is determined exclusively by the unperturbed photonic crystal.

Next, we simplify Eq.\eqref{eq:kp_Hamiltonian_2x2}
by using the mirror symmetry along the $x$ axis,
which is independent of the glide symmetry.
We set the unit cell as shown in Fig. \ref{fig_2Dlossy} (a),
and $x$-axis as the center of the unit cell.
Since the initial permittivity and perturbation are symmetric about the $x$ axis,
$\varepsilon(\bm{r})$ and $\Delta \varepsilon(\bm{r})$ in the unit cell are invariant 
under the mirror operation with respect to the $yz$-plane:
\begin{align}
  \label{eq:eps_mirror_symmetry}
  \hat{M}_x \left( \varepsilon(\bm{r}) \right)
  &= \varepsilon(-x,y,z) = \varepsilon(x,y,z)\\
  \label{eq:deps_mirror_symmetry}
  \hat{M}_x \left( \Delta\varepsilon(\bm{r}) \right)
  &= \Delta\varepsilon(-x,y,z) =\Delta\varepsilon(x,y,z)
\end{align}
Here, $\hat{M}_x$ is the mirror operator with respect to the $yz$-plane.
Note that the assumption that the initial permittivity are symmetric 
about the $x$-axis (Eq.\eqref{eq:deps_mirror_symmetry})
is generally satisfied in typical photonic crystal lattice such as triangular,
square, and honeycomb lattices.
%
%Since $\varepsilon(\bm{r})$ has the mirror symmetry along the $x$ axis, 
Then, the two eigenmodes can be classified into the following two categories 
according to the eigenvalues for operator $\hat{M}_x$:
\begin{align}
    \hat{M}_x {\left( \bm{\varphi}_{ik_0}^{(0)}(x,y,z) \right)}_z
    &= {\left( \bm{\varphi}_{ik_0}^{(0)} \left(-x,y,z\right) \right)}_z \nonumber\\
    &= \pm {\left(\bm{\varphi}_{ik_0}^{(0)} (x,y,z) \right)}_z, \quad i\in \{ \mathrm{u}, \mathrm{d} \}
\end{align}
There is an arbitrary choice of the basis functions of the eigenmodes at $k_x = \pi/a$
since the band dispersion curves are degenerate at the Dirac point,
and we choose the basis functions 
$\bm{\varphi}_{\mathrm{u} k_0}^{(0)}(\bm{r}), \bm{\varphi}_{\mathrm{d} k_0}^{(0)}(\bm{r})$,
which are characterized by the mirror eigenvalues $-1$ and $1$:
\begin{align}
  \label{eq:kp_basis_u_from_mirror}
  \hat{M}_x {\left( \bm{\varphi}_{\mathrm{u} k_0}^{(0)}(\bm{r}) \right)}_z
  &= -{\left( \bm{\varphi}_{\mathrm{u} k_0}^{(0)}(\bm{r}) \right)}_z \\
  \label{eq:kp_basis_d_from_mirror}
  \hat{M}_x {\left( \bm{\varphi}_{\mathrm{d} k_0}^{(0)}(\bm{r}) \right)}_z
  &= {\left( \bm{\varphi}_{\mathrm{d} k_0}^{(0)}(\bm{r}) \right)}_z
\end{align}
From the relationship of electric and magnetic fields of the eigenmodes,
The symmetry of $H_z, E_x, E_y$ at the Dirac point is summarized in Table \ref{tab:table4}.
\begin{table}
  \caption{\label{tab:table4}%
  The symmetry along the $y$-axis of the eigenmodes at the Dirac point.
  The value of 1 or $-1$ represent the symmetric and antisymmetric, respectively.
  }
  \begin{ruledtabular}
    \begin{tabular}{ccc}
    &u&d\\
    \hline
    $H_z$ &$-1$&$1$\\
    $E_x$ &$-1$&$1$\\
    $E_y$ &$1$&$-1$\\
    \end{tabular}
  \end{ruledtabular}
\end{table}
According to the even or odd symmetry of the electromagnetic field components
as shown in Table \ref{tab:table4},
\begin{align}
  \label{eq:zero_condition}
  P_{\mathrm{u}\mathrm{u}} = P_{\mathrm{d}\mathrm{d}} =
  G_{\mathrm{u}\mathrm{d}} = G_{\mathrm{d}\mathrm{u}} = Q_{\mathrm{u}\mathrm{d}} = Q_{\mathrm{d}\mathrm{u}} = 0
\end{align}
holds
%$P_{\mathrm{u}\mathrm{u}}, P_{\mathrm{d}\mathrm{d}}$, $G_{\mathrm{u}\mathrm{d}}$, $G_{\mathrm{d}\mathrm{u}}$, $Q_{\mathrm{u}\mathrm{d}}$, $Q_{\mathrm{d}\mathrm{u}}$ in Eq.\eqref{eq:kp_Hamiltonian_2x2} are zero
when the condition \eqref{eq:eps_mirror_symmetry} and \eqref{eq:deps_mirror_symmetry} are satisfied.
Moreover, we get 
${\left( P_{\mathrm{u} \mathrm{d}} \right)}_x = -{\left( P_{\mathrm{d} \mathrm{u}} \right)}_x =:-iP$,
$P \in \mathbf{R}$ from the time reversal symmetry of the initial structure, 
and $Q_{\mathrm{u} \mathrm{u}} = Q_{\mathrm{d} \mathrm{d}} =:Q$ from the mirror symmetry.
Therefore, the Eq. \eqref{eq:kp_Hamiltonian_2x2} is reduced to the following form:
\begin{align}
  &\left[
    \begin{matrix}
      F & iP \delta k_x \\
      -iP \delta k_x & F
    \end{matrix}
    \right]
    \left[
    \begin{matrix}
      C_\mathrm{u} \\
      C_\mathrm{d}
    \end{matrix}
    \right] \nonumber\\ &\qquad =
    {\left( \frac{\omega_{nk}}{c} \right)}^2
    \left[
    \begin{matrix}
      1 + G_{\mathrm{u}\mathrm{u}} & 0 \\
      0 & 1 + G_{\mathrm{d}\mathrm{d}}
    \end{matrix}
    \right]
    \left[
    \begin{matrix}
      C_\mathrm{u} \\
      C_\mathrm{d}
    \end{matrix}
    \right], \\
    &\qquad F:={\left( \frac{\omega_D}{c} \right)}^2 + Q {(\delta k_x)}^2
\end{align}
Here,
$\omega_D$ is the eigenfrequency of two eigenmodes at the Dirac point, 
and $P$ corresponds to the slope of the band dispersion curves at $k_x a=\pi$.
Furthermore, multiply the inverse of the diagonal matrix from the left 
to eliminate the matrix from the right side of the equation, 
and let 
\begin{align}
  \label{eq:def_DeltaG}
  \Delta G &:= 
  \frac{ G_{\mathrm{d}\mathrm{d}} - G_{\mathrm{u}\mathrm{u}} }{2}
  \nonumber\\
  &=
  \frac{1}{2}
  \int_{\mathrm{u.c.}} \mathrm{d}^3r 
  \Delta \varepsilon (\bm{r}) 
  \left( 
    {\left|
      \bm{\psi}_{\mathrm{d} k_0 }^{(0)}(\bm{r}) 
    \right| }^2- {\left|
      \bm{\psi}_{\mathrm{u} k_0}^{(0)}(\bm{r})
    \right| }^2
  \right)
\end{align}
% $\Delta G := \left( G_{\mathrm{d}\mathrm{d}} - G_{\mathrm{u}\mathrm{u}} \right) / 2$,
then we get the following form.
\begin{align}
  \label{eq:kp_glide2x2}
  &\left(F 
    \left( 
      1 + \frac{ G_{\mathrm{u}\mathrm{u}} + G_{\mathrm{d}\mathrm{d}}}{2} 
    \right) I_2 + H' 
  \right)
  \left[
  \begin{matrix}
    C_\mathrm{u} \\
    C_\mathrm{d}
  \end{matrix}
  \right] = E_{nk}
  \left[
  \begin{matrix}
    C_\mathrm{u} \\
    C_\mathrm{d}
  \end{matrix}
  \right], \ 
  \nonumber\\
  &\quad H' =
  \left[
    \begin{matrix}
      F \Delta G & i(1+G_{\mathrm{d}\mathrm{d}})P \delta k_x \\
      -i(1+G_{\mathrm{u}\mathrm{u}})P \delta k_x & -F \Delta G
    \end{matrix}
    \right]
\end{align}
Here, $I_{2}$ is a $2\times 2$ identity matrix, and
$\displaystyle E_{nk}:= (1 + G_{\mathrm{u}\mathrm{u}}) (1 + G_{\mathrm{d}\mathrm{d}}) {\left( \omega_{nk}/c \right)}^2$
denotes the eigenvalues.
$H'$ in Eq.\eqref{eq:kp_glide2x2} characterizes the band structure of the edge states.
The diagonal term $\pm {\left( \omega_D / c \right)}^2 \Delta G$ in Eq. \eqref{eq:kp_glide2x2} 
is regarded as the difference between the potentials of $y>0$ and $y<0$. 
More specifically, 
$\Delta G$ is proportional to the difference between two norm of the basis functions in the perturbed region.
Furthermore, the off-diagonal terms 
$i(1+G_{\mathrm{d}\mathrm{d}})P,\ -i(1+G_{\mathrm{u}\mathrm{u}})P$ in $H'$ 
represent the complex couplings of two basis.
Note that it is fundamentally important for the simplicity of the Hamiltonian thanks to Eq.\eqref{eq:zero_condition} that the basis $\mathrm{u}$ and $\mathrm{d}$ are degenerated due to the glide symmetry, and that they also possess (anti-)symmetry with respect to $\hat{M}_x$.

The corresponding eigenvalues $E_{nk}$ are given by the following equation:
\begin{align}
  \label{eq:kp_eigenvalue}
  E_{nk} &= 
  F
  \left(
    1 + \frac{ G_{\mathrm{u}\mathrm{u}} + G_{\mathrm{d}\mathrm{d}}}{2} 
  \right)
  \nonumber\\
  &\ \pm \sqrt{ 
    {(F\Delta G)}^2 
    + 
    (1 + G_{\mathrm{u}\mathrm{u}}) (1 + G_{\mathrm{d}\mathrm{d}})
    {\left( P \, \delta k_x \right)}^2 
  }
\end{align}
The sign of the square root in Eq. \eqref{eq:kp_eigenvalue}
corresponds to the solution number of the eigenvalues $n=1,2$.
From the form of eigenvalues in Eq. \eqref{eq:kp_eigenvalue},
The corresponding EP condition is given by
\begin{align}
  \label{eq:kp_ep_condition}
  \frac{{P(\delta k_x)}^2}{ {\left( {\left( \frac{\omega_D}{c} \right)}^2 + Q {(\delta k_x)}^2 \right)}^2 } =
  \frac{-{(\Delta G)}^2}{ (1+G_{\mathrm{u}\mathrm{u}})(1+G_{\mathrm{d}\mathrm{d}}) }
\end{align}
All of the components in the left-hand side of Eq. \eqref{eq:kp_ep_condition} are real numbers,
there exists EPs in the real $\delta k_x$-space 
when the right-hand side of Eq. \eqref{eq:kp_ep_condition} is positive.
The right-hand side of Eq. \eqref{eq:kp_ep_condition} is always positive
as long as the system has the glide and time reversal symmetry,
since $\Delta G$ is pure imaginary
and the $(1 + G_{\mathrm{u}\mathrm{u}}) (1 + G_{\mathrm{d}\mathrm{d}})$ is a real number
because $G_{\mathrm{u}\mathrm{u}}^* = G_{\mathrm{d}\mathrm{d}}$ holds.
Eq. \eqref{eq:kp_ep_condition} indicates that the EPs under the glide and time reversal symmetry
always exist near $k_x = \pi/a$,
which means that EP is always located below the light line.

Although the initial form of Hamiltonian shown in Eq.\eqref{eq:kp_Hamiltonian_2x2}
has the similar form in Ref. \cite{PhysRevLett.116.203902},
the derived Hamiltonian shown in Eq.\eqref{eq:kp_glide2x2} 
and the physical interpretation are entirely different.
Their model targets the bulk bands in the photonic crystals, 
and utilizes the band folding with double-period imaginary perturbations.
The band folding places the EP above the light line, 
which makes it difficult to use as a waveguide mode.
In contrast, our system is a photonic crystal waveguide with edge states,
which allows the EP to utilize the waveguide mode.
Furthermore, the design with glide symmetry allows the EP to be located below the light line, 
which ensures that the EP modes are free from out-of-plane radiation loss.
This aspect highlights an advantage of our scheme 
compared to the waveguide structures proposed in Refs. \cite{PhysRevB.100.115412, Fang_2019, Mock:20}.

%\subsection{Numerical verification of $\bm{k} \cdot \bm{p}$ Hamiltonian}

%%%%% Figure 4 %%%%%
\begin{figure}
  \includegraphics[width=7.5cm]{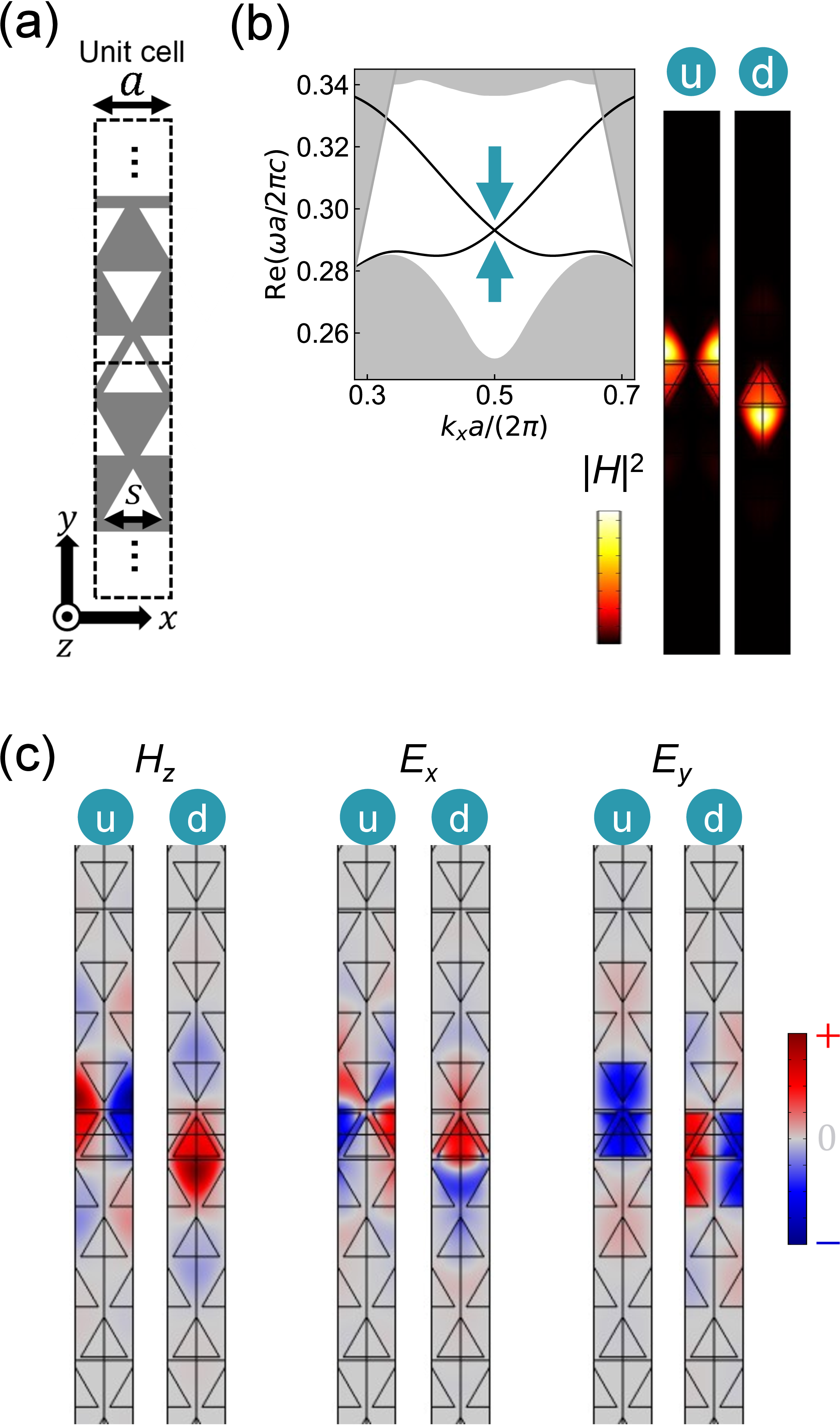}
  \caption{\label{fig_2Dkp}
      (a) The configuration of the photonic crystal heterostructure 
          to obtain the eigenmodes at the Dirac point used for the $\bm{k} \cdot \bm{p}$ Hamiltonian.
      (b) ${|\bm{H}|}^2$ distribution of the eigenmodes in the photonic crystal heterostructure.
      Here, label u, d denote $\bm{\varphi}_{\mathrm{u}k_0}^{(0)}, \bm{\varphi}_{\mathrm{d}k_0}^{(0)}$, respectively.
      (c) $H_z$, $E_x$, and $E_y$ distribution of the bases at the Dirac point.
  }
\end{figure}
The configuration for numerical analysis
and the comparison between FEM and $\bm{k} \cdot \bm{p}$ approximation are as follows.
Figures \ref{fig_2Dkp}(a) and (b) show the computational setup 
and the magnetic field of the two bases, respectively.
We can see that
${| \bm{\varphi}_\mathrm{u}( \bm{r} ) |}^2$ 
and ${| \bm{\varphi}_\mathrm{d}( \bm{r} ) |}^2$,
which are proportional to ${|\bm{H}|}^2$,
%which are proportional to ${| \bm{H}_\mathrm{u}( \bm{r} ) |}^2$ and ${| \bm{H}_\mathrm{d}( \bm{r} ) |}^2$,
are localized at the upper and lower sites, respectively.
Figure \ref{fig_2Dkp} (c) shows the $H_z$, $E_x$, and $E_y$ distribution 
of the bases at the Dirac point 
defined by Eq. \eqref{eq:kp_basis_u_from_mirror} and \eqref{eq:kp_basis_d_from_mirror}.
It is clear the $H_z$ and $E_x$ distribution of mode u are antisymmetric along the $y$-axis,
while the $E_y$ distribution is symmetric along the $y$-axis.
Moreover, the opposite is true for mode d.
These results are consistent with the classification of the eigenmodes
based on the mirror symmetry along the $x$ axis in Table \ref{tab:table4}.

%%%%% Figure 4 %%%%%
\begin{figure*}
  \includegraphics[width=18cm]{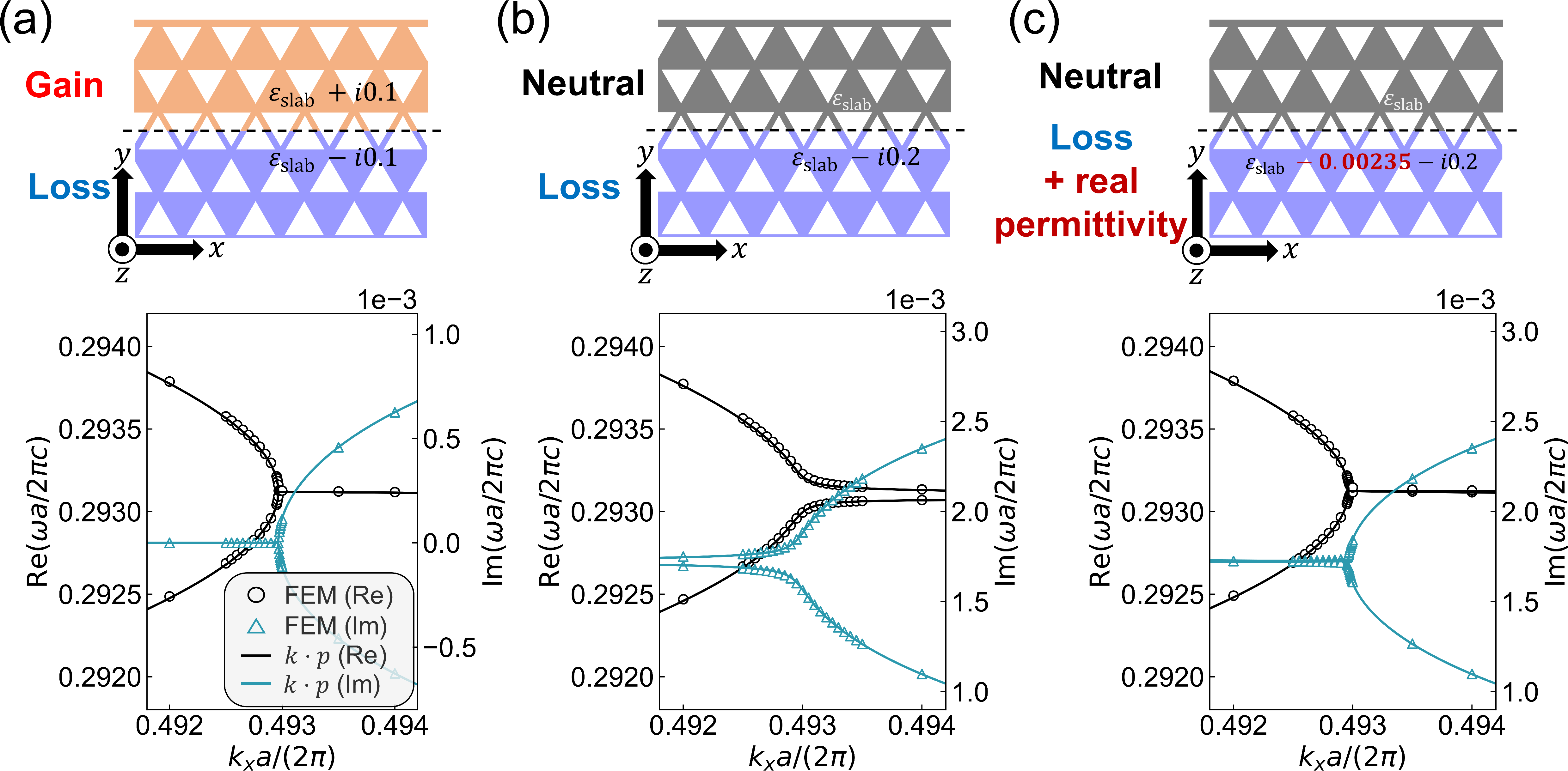}
  \caption{\label{fig_2DkpvsFEM}
      Comparison of the band dispersion with the FEM calculation and the $\bm{k} \cdot \bm{p}$ Hamiltonian. 
      The black circle and blue triangle points represent the real and imaginary parts of the eigenvalues 
      obtained by FEM, while the black and blue lines represent the real and imaginary parts 
      by the $\bm{k} \cdot \bm{p}$ Hamiltonian.
      (a) PT-symmetric permittivity perturbation 
      with $\Delta \varepsilon = i0.1$ for $y > 0$ and 
      $\Delta \varepsilon = -i0.1$ for $y < 0$.
      (b) $\Delta \varepsilon = -i0.2$ only for $y < 0$.
      (c) $\Delta \varepsilon = -0.00235 - i0.2$ only for $y < 0$.
      }
\end{figure*}
The band dispersion in a heterostructure with the PT-symmetric permittivity perturbation
using the FEM method and $\bm{k} \cdot \bm{p}$ Hamiltonian
is shown in Fig. \ref{fig_2DkpvsFEM} (a).
The black circle and blue triangle points represent the real and imaginary parts of the eigenvalues 
obtained by FEM, and the black and blue lines represent the real and imaginary parts 
by the $\bm{k} \cdot \bm{p}$ Hamiltonian.
No fitting is used to obtain the coefficients in the $\bm{k} \cdot \bm{p}$ Hamiltonian;
instead, the coefficients are determined using two unperturbed bases 
$\bm{\varphi}_{\alpha k_0}^{(0)}(\bm{r})$ and $\bm{\psi}_{\alpha k_0}^{(0)}(\bm{r})$
calculated by FEM and Eq. \eqref{eq:def_matG}, \eqref{eq:def_matP} and \eqref{eq:def_matQ}.
The band dispersion show the good agreement between the $\bm{k} \cdot \bm{p}$ and FEM methods.
Moreover, the EPs are clearly observed in the real $\delta k_x$-space,
which is consistent with the theoretical prediction.

\subsection{Restoration of EP by adjusting the real part of the permittivity}
\label{sec:restoration_EP}

Next, we describe how to restore the EP in the case %by adjusting the real part of the permittivity
where the EP is smoothed out by breaking the glide and time reversal symmetry.
It is obvious that 
if there is a mismatch in the real part of the permittivity on both sides of the glide plane, 
as shown in Fig. \ref{fig_2Dlossy}(a),
the EP can be restored by simply adjusting the real part of the permittivity 
that compensates for the difference.
Mathematically, this can be explained as follows: 
when there is a mismatch in the real part of the permittivity, 
$\Delta G$ defined in Eq. \eqref{eq:def_DeltaG}
becomes a complex number, and the EP condition shown in Eq. \eqref{eq:kp_ep_condition}
is no longer satisfied. 
By compensating for the real part of the permittivity, 
$\Delta G$ becomes a pure imaginary number again, and the EP is restored.
In contrast to the situation, 
the way to restore the EP in the presence of gain-loss imbalance is not trivial.
The gain-loss imbalance breaks the EP condition since 
$G_{\mathrm{u}\mathrm{u}}^* \neq G_{\mathrm{d}\mathrm{d}}$ holds and
$(1 + G_{\mathrm{u}\mathrm{u}}) (1 + G_{\mathrm{d}\mathrm{d}})$
becomes a complex number.
The situation has the same mathematical origin 
as the smoothing of EP due to the complex coupling between the two basis 
in a coupled resonator array \cite{PhysRevA.105.013523}.

The following discussion will focus on the case 
where the uniform imaginary part of the permittivity $\Delta \varepsilon_i$ is introduced 
for the dielectric region where $y < 0$
($V_G$ as shown in the blue region in Fig. \ref{fig_2Dwg} (b),(c)),
which is responsible for the non-Hermitian perturbation.
Moreover, the real part of the permittivity perturbation $\Delta \varepsilon_r$ 
is introduced for $V_G$ to restore the EP.

Let $I_{\alpha \beta}$, $J_{\alpha \beta}$, and $\kappa$ be defined as follows:
\begin{align}
  I_{\alpha \beta} &:=
  \int_{\bm{r} \in V_G} 
  \bm{\psi}_{\alpha k_0}^{(0)*}(\bm{r}) \cdot \bm{\psi}_{\beta k_0}^{(0)}(\bm{r})
  \mathrm{d}^3r, 
  \nonumber\\
  J_{\alpha \beta} &:= \Delta \varepsilon_\mathrm{i} I_{\alpha \beta}, \quad
  \kappa := \frac{\Delta \varepsilon_\mathrm{r}}{\Delta \varepsilon_\mathrm{i}}
\end{align}
From the following definition, we get $G_{\alpha \beta}= (\kappa + i)J_{\alpha \beta}$.
The necessary condition 
for $-{(\Delta G)}^2 / (1+G_{\mathrm{u}\mathrm{u}})(1+G_{\mathrm{d}\mathrm{d}})$ 
to be a positive real number,
which is the condition for the existence of EP shown in Eq. \eqref{eq:kp_ep_condition},
is that the vectors of the denominator and numerator are parallel in the complex plane, 
so the following conditional expression is obtained.
\begin{align}
  \label{eq:ep_condition_matG}
  \mathrm{det}
  \left[
  \begin{matrix}
    \mathrm{Re}\left( -{(\Delta G)}^2 \right) & \mathrm{Re}\left( (1+G_{\mathrm{u}\mathrm{u}})(1+G_{\mathrm{d}\mathrm{d}}) \right) \\
    \mathrm{Im}\left( -{(\Delta G)}^2 \right) & \mathrm{Im}\left( (1+G_{\mathrm{u}\mathrm{u}})(1+G_{\mathrm{d}\mathrm{d}}) \right)
  \end{matrix}
  \right]
  &= 0
\end{align}
Substitute $\Delta G, \ G_{\mathrm{u}\mathrm{u}}$ and $G_{\mathrm{d}\mathrm{d}}$ 
into $\kappa$ and $J_{ij}$, 
and then we get the relation under EP conditions:
\begin{align}
  \label{eq:kp_ep_condition_kappa}
   (J_{\mathrm{u}\mathrm{u}} + J_{\mathrm{d}\mathrm{d}})\kappa^2 
   + 2\kappa + (J_{\mathrm{u}\mathrm{u}} + J_{\mathrm{d}\mathrm{d}}) = 0
\end{align}
By solving Eq. \eqref{eq:kp_ep_condition_kappa} for $\kappa$
and choose a physical solution $\kappa \ll 1$,
then we obtain the relationship 
for $\Delta \varepsilon_\mathrm{r}$ and $\Delta \varepsilon_\mathrm{i}$ at EP, 
which is the condition we are looking for.
\begin{align}
  \label{eq:kp_ep_permittivity}
  \Delta \varepsilon_\mathrm{r} 
  = - \frac{1 - \sqrt{1-(I_{\mathrm{u}\mathrm{u}} + I_{\mathrm{d}\mathrm{d}})^2{(\Delta \varepsilon_\mathrm{i})}^2 }}{ I_{\mathrm{u}\mathrm{u}} + I_{\mathrm{d}\mathrm{d}} }
\end{align}
Here, we numerically verify the EP smoothing and restoration 
by adjusting the real part of the permittivity.
Figure \ref{fig_2DkpvsFEM}(b) shows the band dispersion 
when the imaginary part of the permittivity is $\Delta \varepsilon_i(r) = -0.2$ only for $y < 0$. 
It reveals that the EPs are smoothed out unlike the case in Fig. \ref{fig_2DkpvsFEM} (a).
In this case, $(1 + G_{\mathrm{u}\mathrm{u}}) (1 + G_{\mathrm{d}\mathrm{d}})$ is a complex number 
and EP does not exist in real $\delta k_x$-space.
According to Eq. \eqref{eq:kp_ep_permittivity}, 
$\Delta \varepsilon_\mathrm{r} = -0.00235$ for $y<0$ is required 
to cancel the effects of complex coupling 
in the case of Fig. \ref{fig_2DkpvsFEM}(b) with $\Delta \varepsilon_\mathrm{i} = -0.2$ for $y<0$.
Figure \ref{fig_2DkpvsFEM}(c) shows the band dispersion 
when $\Delta \varepsilon_\mathrm{r} = -0.00235$ for $y<0$ is introduced,
and the EP smoothing is clearly suppressed compared to Fig. \ref{fig_2DkpvsFEM}(b).
Thus, even when the EP is smoothed out by the gain-loss imbalance, 
appropriate adjustment of the real part of the permittivity offsets the complex coupling
and restores the anomalous dispersion.
In this way, even for modes in photonic crystal waveguides, where a formalism like tight-binding is not well known, the effective Hamiltonian model can elucidate the conditions for EP restoration in non-trivial gain-loss imbalance scenarios.

\section{Superluminal edge states and enlarged propagation length on photonic crystal waveguide}

%%%%% Figure 5 %%%%%
\begin{figure}
  \includegraphics[width=7cm]{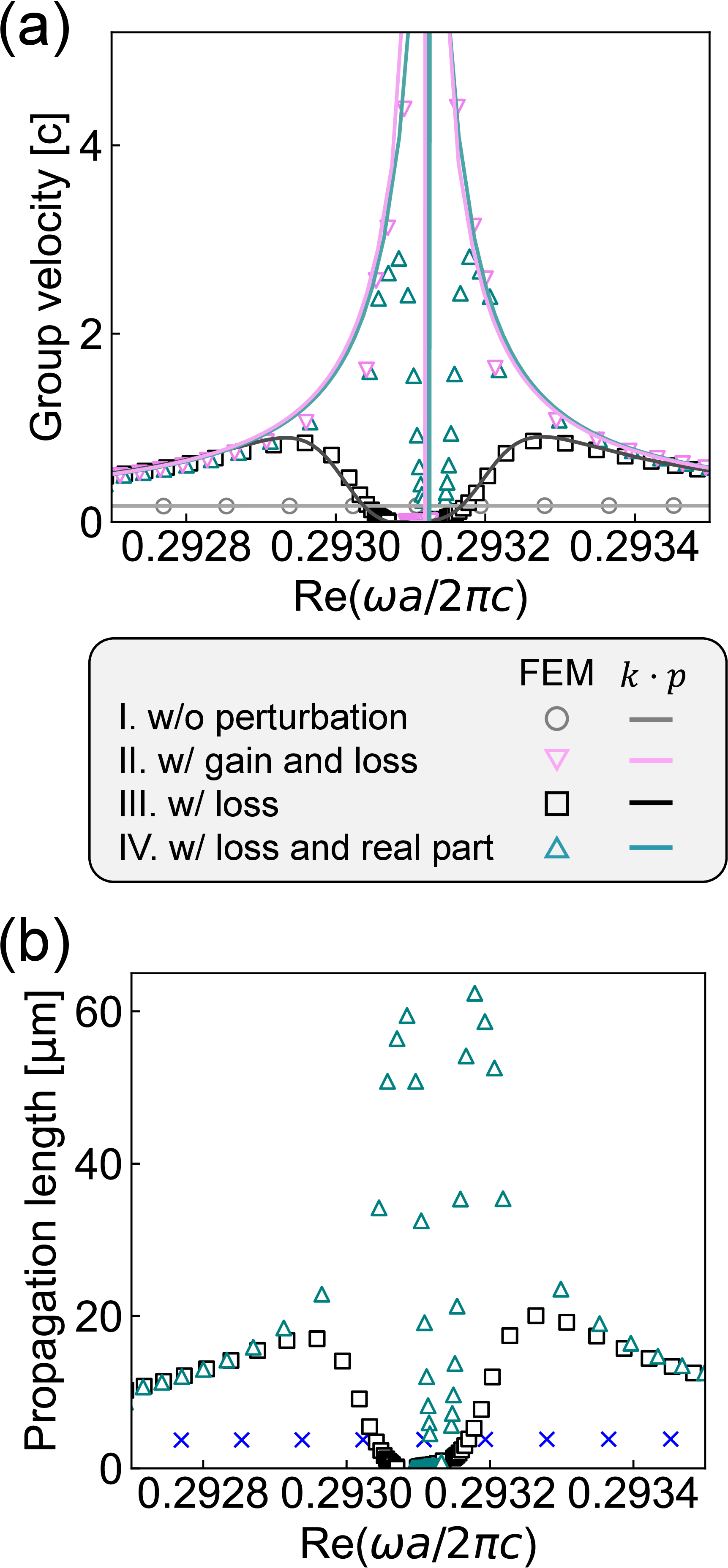}
  \caption{\label{fig_2Dvg}
    (a) Group velocity of the edge states in the photonic crystal heterostructure
    I. without permittivity adjustment, 
    II. with $\Delta \varepsilon = i0.1$ for $y > 0$ and $\Delta \varepsilon = -i0.1$ for $y < 0$ (balanced loss-gain),
    III. with $\Delta \varepsilon = -i0.2$ for $y > 0$,
    IV. with $\Delta \varepsilon = -0.00235-i0.2$ for $y < 0$.
    II., III. and IV. represent the cases in Fig. \ref{fig_2DkpvsFEM} (a), (b) and (c), respectively.
    (b) Propagation length of the edge states in the FEM calculation in III. and IV.
    The blue cross denotes with initial permittivity and the same lifetime of the edge states 
    at $k_x a /(2\pi) = 0.49$ in (a) III.
  }
\end{figure}
The anomalous dispersion near the EPs leads to the high group velocity of the edge states,
which allows the realization of superluminal propagation.
Here, we numerically verify the group velocity of the edge states, 
and show that the singularity in the band dispersion near the EP is restored 
by the method we proposed in the previous section, 
resulting in a group velocity exceeding the light speed in vacuum $c$. 
In addition, we discuss the fast-light effect on the propagation length of the edge states.
The propagation length of the edge mode is limited by the lifetime of the edge states in the passive systems, 
but is expected to increase as the group velocity increases. 
In the following sections, we will also discuss the expected increase in propagation length 
based on the anomalous group velocity near the EPs.

Figure \ref{fig_2Dvg} (a) shows the group velocity of the edge states,
which is calculated by the slope of the band dispersion in Fig. \ref{fig_2DkpvsFEM}.
The group velocity is defined as
$v_g = \partial \omega_R / \partial k_x$, 
where $\omega_R$ is the real part of the eigenfrequency.
\begin{enumerate}
\renewcommand{\labelenumi}{\Roman{enumi}.}
  \item without permittivity adjustment
  \item with $\Delta \varepsilon = i0.1$ for $y > 0$ and $\Delta \varepsilon = -i0.1$ for $y < 0$ (balanced loss-gain)
  \item with $\Delta \varepsilon = -i0.2$ for $y > 0$
  \item with $\Delta \varepsilon = -0.00235-i0.2$ for $y < 0$
\end{enumerate}
Here, II., III. and IV. represent the cases in Fig. \ref{fig_2DkpvsFEM} (a), (b) and (c), respectively.
When the permittivity is not adjusted (I.),
the maximum group velocity of the edge states is $v_g = 0.17c$.
Note that the initial group velocity is larger than that of CROW ($v_g = 0.02c$ in Ref \cite{PhysRevA.105.013523}).
Then, the group velocity is increased by the balanced loss-gain perturbation $\Delta \varepsilon = \pm i0.1$ (II.).
The maximum group velocity of the edge states reaches the superluminal ($v_g = 4.5c$),
which is determined by the finite mesh of the wavevector $\Delta k_x$ in the FEM calculation
and should theoretically diverge at the EPs.
The graph III. in Fig. \ref{fig_2DkpvsFEM} (a) shows the group velocity of the edge states 
with loss-biased perturbation $\Delta \varepsilon = -i0.2$ for $y > 0$.
The maximum group velocity near the EPs in FEM calculation is subluminal ($v_g = 0.85c$),
which is 5.0 times larger than that of $\Delta \varepsilon = 0$ ($v_g = 0.17c$),
but is limited by the EP smoothing.
%Note that the $\bm{k} \cdot \bm{p}$ Hamiltonian reproduces the FEM calculation well.
%
The graph IV. in Fig. \ref{fig_2Dvg} (a) shows the group velocity of the edge states
with the permittivity adjustment $\Delta \varepsilon = -0.00235-i0.2$ for $y < 0$.
The adjustment of the real part of the permittivity restore the EP,
and the maximum group velocity of the edge states is accelerated to $v_g = 2.8c$,
which is 16.9 times larger than that of $\Delta \varepsilon = 0$.
The group velocity calculated from the FEM method is sufficiently close to the singularity 
predicted by the $\bm{k} \cdot \bm{p}$ method.
Consequently, by restoring the EP through adjustment of the real part of the permittivity, the group velocity can be increased from subluminal to superluminal.

Since the photon lifetime determined by the absorption is independent of the group velocity,
the propagation length of the edge states is expected to increase
as the group velocity increases.
The propagation length of the edge state $l$ is defined as the length 
at which the intensity of the light decreases to $1/e$:
\begin{align}
  \label{eq:def_propagation_length}
  l = \frac{\partial \omega_R}{\partial k_x} \frac{1}{2\omega_I},
\end{align}
Here, $\omega_R$ and $\omega_I$ are the real and imaginary parts of the eigenfrequency.
Figure \ref{fig_2Dvg} (b) shows the propagation length of the edge states.
%The gray and blue points represent the results with Fig. \ref{fig_2DkpvsFEM} (a) and (b), respectively.
Here, the case with balanced loss-gain perturbation (Fig. \ref{fig_2DkpvsFEM} (a)) is not shown
since the propagation length is infinite due to the negative imaginary part of the eigenfrequency.
The maximum propagation length is 20 $\mathrm{\mu}$m 
when the permittivity adjustment $\Delta \varepsilon = -i0.2$ for $y > 0$ is introduced.
Moreover, the propagation length reaches 62 $\mathrm{\mu}$m
with the real part of permittivity adjustment $\Delta \varepsilon = -0.00235-i0.2$ for $y < 0$,
which is due to the suppression of the EP smoothing.
The enlarged propagation length despite short lifetime ($\tau := 1/(2\omega_I) = 0.072$ ps)
is achieved by the high group velocity effect near the EPs.
Assuming the same group velocity in the unperturbed case 
and the same lifetime at EP for the perturbed cases (IV. in Fig. \ref{fig_2Dvg} (a)), 
the propagation length is limited to 4.0 $\mathrm{\mu}$m. 
In this way, the propagation length of the edge states can be enlarged even in the lossy system, 
which may allow us to utilize the guided mode even in the passive system.

\section{implementation into the graphene-loaded photonic crystal slab}
\label{sec:graphene}

Our scheme for generating and restoring EPs in photonic crystal waveguides 
can be applied to the realistic silicon photonic crystal slabs.
In this section, we show how to implement our scheme into the graphene-loaded photonic crystal slab.

%%%%% Figure 6 %%%%%
\begin{figure*}
  \includegraphics[width=17.5cm]{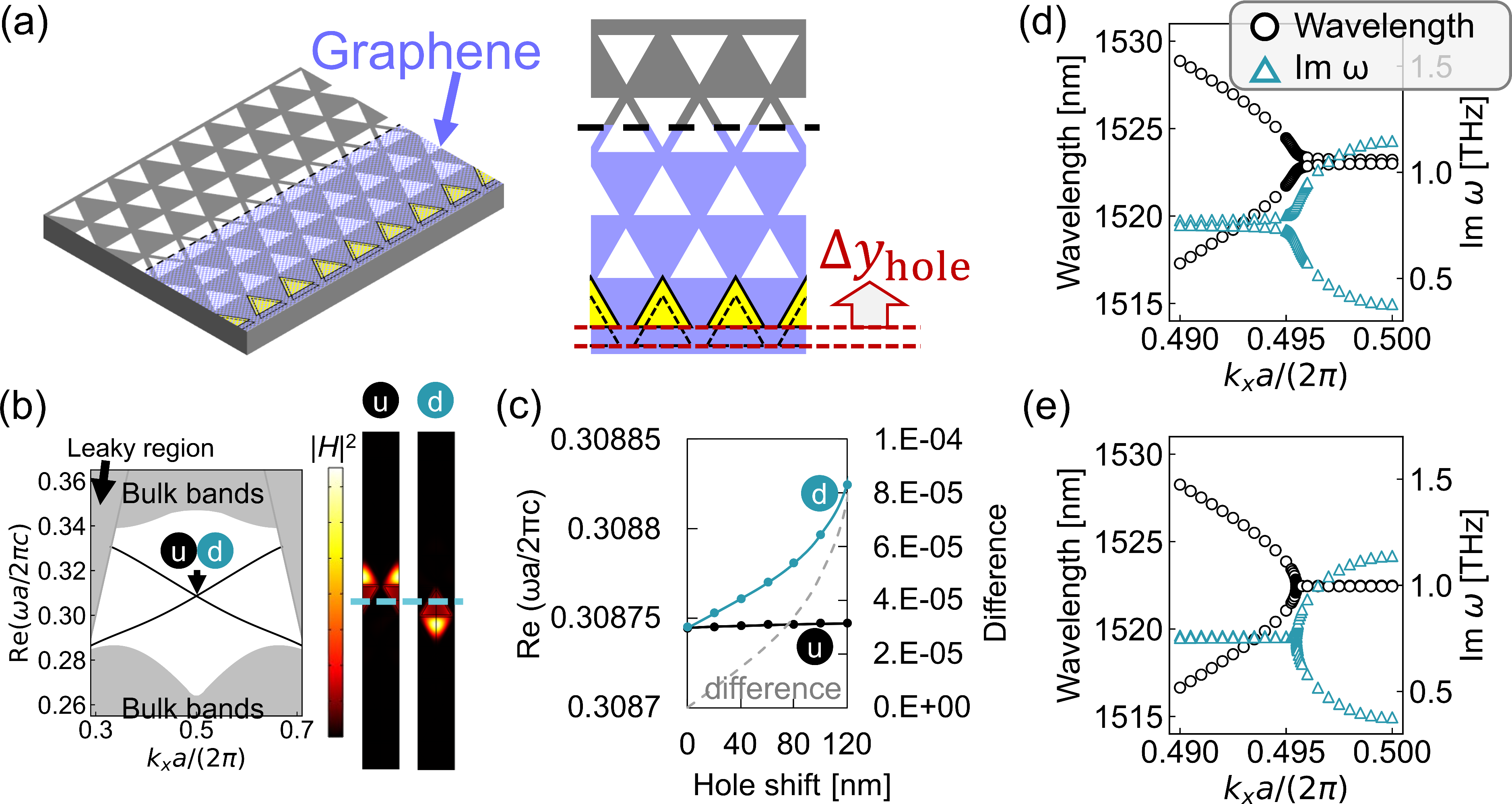}
  \caption{\label{fig_3Dfem}
  (a) Schematic of the photonic crystal slab loaded with 5 sheet of graphene.
  The right panel describes the hole shift of the position of the triangular hole $\Delta y_\mathrm{hole}$.
  (b) The band dispersion of the graphene-loaded photonic crystal slab
  and ${|\bm{H}|}^2$ distribution of the eigenmodes at the initial Dirac point.
  (c) The frequency of the two eigenmodes at $k_x = \pi/a$ without graphene
  as a function of the triangular hole shift $\Delta y_\mathrm{hole}$.
  (d), (e) The band dispersion of the graphene-loaded photonic crystal slab
  (d) without no hole shift and (e) with the hole shift $\Delta y_\mathrm{hole} = 104$ nm.
  }
\end{figure*}
Figure \ref{fig_3Dfem} (a) shows a schematic 
of the graphene loaded photonic crystal slab.
We assume the lattice constant is $a=470$ nm, the one side of triangular hole is $s=0.8a$,
which is the same as the previous 2D simulations in Fig. \ref{fig_2Dwg} (a).
The thickness of the slab is set to $h=200$ nm, 
and the permittivity of the slab is set to $\varepsilon_\mathrm{slab} = {(3.48)}^2$.
Then, the 5-sheet of graphene is placed 
on the top of the slab and one side of the heterostructures ($y<0$) to generate the EPs.
Graphene has remarkable properties for applications in optical devices,
such as large absorption in one place, reconfigurable conductivity \cite{RevModPhys.81.109}. 
It is well known that the optical absorption of graphene depends on its Fermi level, 
which can be controlled by voltage injection
\cite{PhysRevLett.99.246803, 10.1063/1.2956669, PhysRevB.83.153410, 
10.1063/1.3291615, Mihnev2016}. 
Our choice of graphene in this study is based solely on its ability 
to easily induce material absorption in photonic crystals,
but the graphene has other advantages, such as the ability to reconfigurably control its permittivity.
Now we aim to purely perturb the imaginary part of the permittivity 
since the large real part of the permittivity breaks the EP condition,
and graphene is particularly well suited for perturbing only the imaginary part of the permittivity,
as it has a large imaginary part of the permittivity in the near-infrared region \cite{10.1063/1.2891452}. 
In addition, some previous reports have demonstrated the successful 
loading and precise patterning of graphene 
onto photonic crystal structures \cite{Majumdar2013, Satoshi:23}, 
further supporting its feasibility for our purposes.
We have assumed that the conductivity of graphene is $\sigma = (6.0882 + i0.1908)\times 10^{-5}$ [S/m] per sheet,
which is the value from the theoretical formula derived from the Kubo formula \cite{10.1063/1.2891452}
with the relaxation time $\tau = 100$ fs, 
the temperature of graphene $T = 300$ K,
the angular frequency of light $\omega = 2\pi c/\lambda$, $\lambda = 1550$ nm,
and the Fermi energy of graphene $E_F=0.2$ eV.
The real and imaginary parts of the conductivity correspond to 
the imaginary and real parts of the permittivity, respectively.
See the supplementary material \ref{sec:appendix_graphene_misalignment} for the effect of the misalignment of the graphene sheet.

Figure \ref{fig_3Dfem} (b) shows the band dispersion of the photonic crystal slab without graphene, 
and we can see that the linear dispersion of edge states appears near $k_x = \pi/a$ 
as in the case in Fig. \ref{fig_2Dwg} (b).
Here, we focus on two eigenmodes at the Dirac point ($k_x = \pi/a$), 
and the modes localized at the top and bottom are named modes u and d, 
as same as the case in Fig. \ref{fig_2Dkp} (b).
The adjustment of the real part of the permittivity 
for one side of the heterostructure ($y<0$) as shown in the right panel of Fig. \ref{fig_2DkpvsFEM} (a)
can be replaced by the position shift of triangular holes.
we choose the triangular hole position in the forth row from the glide symmetry plane
to introduce just the right amount of adjustment,
as shown by the yellow triangles in figure \ref{fig_3Dfem} (a).
Figure \ref{fig_3Dfem} (c) shows the frequency of the two eigenmodes at $k_x = \pi/a$ without graphene
as a function of the $y$-directional triangular hole shift $\Delta y_\mathrm{hole}$.
The frequency of two modes are degenerated in $\Delta y_\mathrm{hole}=0$.
The positive hole shift $\Delta y_\mathrm{hole}$ perturbs only the mode $\mathrm{d}$,
leads to decrease the difference of frequency between two modes
and minimizes near $\Delta y_\mathrm{hole} = 104$ nm.
Therefore, the hole shift of the triangular hole can be used to adjust the effective real part of the permittivity. A details is shown in the supplementary material \ref{sec:appendix_EP_hole_shift}.

Next, we investigate the effect of the graphene loading on the band dispersion.
Figure \ref{fig_3Dfem} (d) shows the band dispersion near $k_x = \pi / a$ 
of the graphene-loaded photonic crystal slab
without the triangular hole shift ($\Delta y_\mathrm{hole} = 0$).
The EPs is smoothed out by the graphene loading,
which is not only due to the imaginary coupling in loss-biased system,
but also due to the positive permittivity modulation by the graphene sheet.
Figure \ref{fig_3Dfem} (e) shows the band dispersion
with the triangular hole shift $\Delta y_\mathrm{hole} = 104$ nm.
The proper adjustment of triangular holes cancels 
the contribution of both the imaginary coupling and the imaginary part of conductivity,
and restores EP.
Consequently, our scheme for generating and restoring EPs in photonic crystal waveguides
can be applied to the silicon photonic crystal slabs
by loading graphene and adjusting the position of triangular holes.

%%%%% Figure 7 %%%%%
\begin{figure}
  \includegraphics[width=7cm]{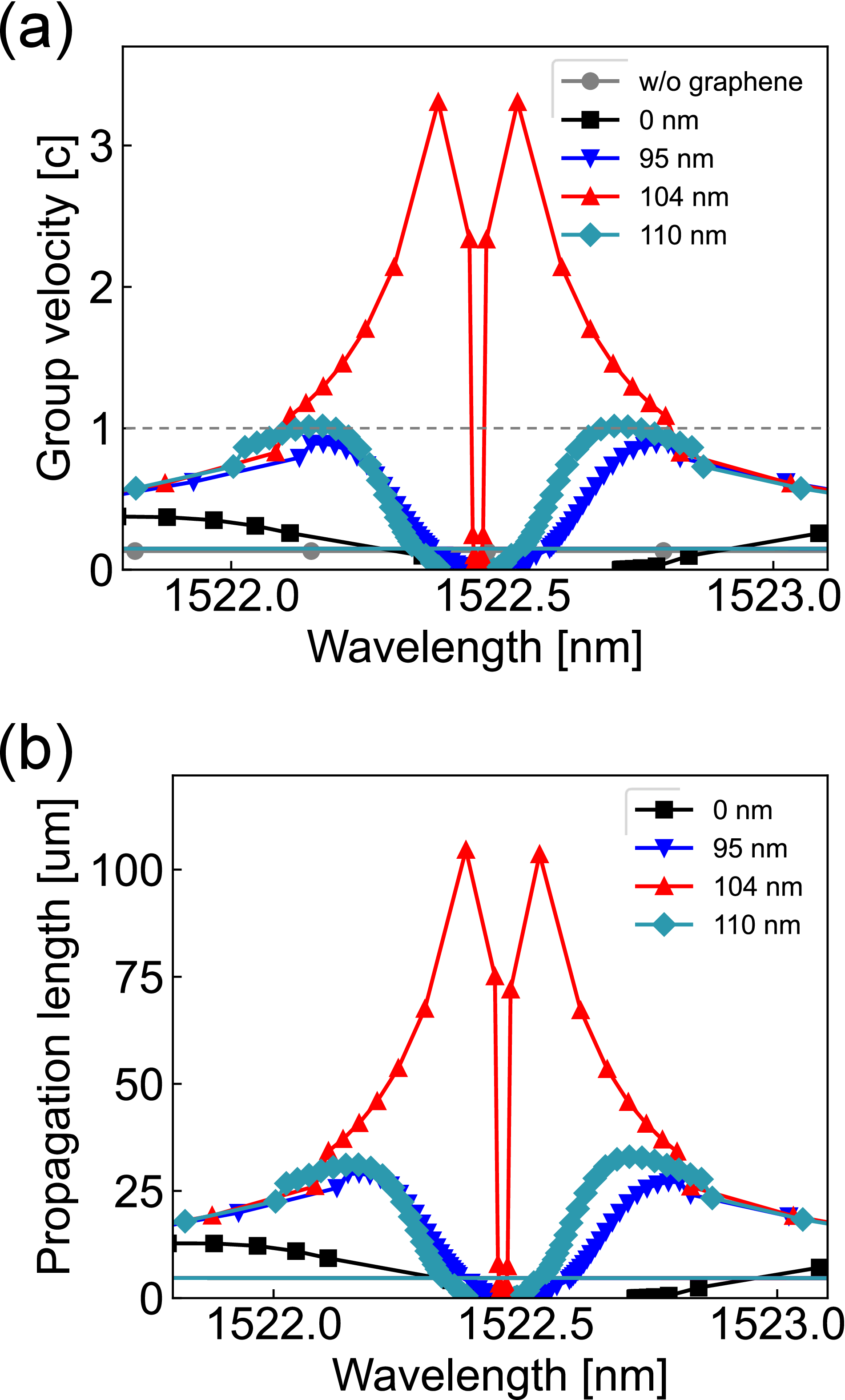}
  \caption{\label{fig_3Dvg}
  (a) Group velocity $v_g$ and (b) Propagation length $l$ of the edge states
  in the graphene-loaded photonic crystal slab near EPs. 
  }
\end{figure}

Finally,
we investigate the group velocity of the edge state and corresponding propagation length
in some triangular hole shift $\Delta y_\mathrm{hole}$.
Figure \ref{fig_3Dvg} (a) shows the group velocity $v_g$ of the edge states
in the graphene-loaded photonic crystal slab near EPs
with some triangular hole shift $\Delta y_\mathrm{hole}$.
The group velocity of initial photonic crystal waveguide near the Dirac cone
is about $v_g = 0.13c$, and increases to $v_g=0.38c$
when graphene is loaded. 
However, without the hole shift, 
the effect of the EP smoothing strongly limits the peak of group velocity.
When a hole shift is added, 
the EP smoothing gradually mitigates and the group velocity peak is maximized 
around $\Delta y_\mathrm{hole}=104$ nm. 
A series of changes in the group velocity with $\Delta y_\mathrm{hole}$ correspond to 
the change of frequency difference at $k_xa=\pi$ between the two modes 
provided by the hole shift,
as shown in Fig. \ref{fig_3Dfem}(d).
The group velocity reaches $v_g = 3.3c$ near the restored EPs
in the case of $\Delta y_\mathrm{hole} = 104$ nm,
and is accelerated to 25 times of that without graphene.
The bandwidth where the group velocity is accelerated over the speed of light in vacuum
is up to 0.4 nm as shown in Fig. \ref{fig_3Dvg} (a),
which corresponds to $1.5$ ps in terms of time width assuming a Gaussian pulse.
Importantly, the propagation length is also elongated by the same ratio, 
directly resulting from the fast light effect.
We confirmed that the propagation length $l$ is elongated to 105 $\mathrm{\mu}$m 
near the restored EPs as shown in Fig. \ref{fig_3Dvg} (b),
which is a staggering number for a lifetime of $0.11$ ps.
It takes 0.38 ps to propagate the same distance in air, 
and it is expected that this difference can be detected for a pulse length of 1.5 ps,
given the ratio of pulse length to group delay 
$(0.38-0.11)/1.5 \approx 0.18$: 
This value is better than the previously reported ratio of pulse width to group delay 
in superluminal propagation experiments 
$(23.8 \mathrm{\mu s})/(0.05 \mathrm{s})\approx 0.048$ \cite{doi:10.1126/science.1084429}. 
In addition, the group velocity of the edge state in a photonic crystal waveguide without graphene is about 1/7 of that in air, 
making it even easier to measure the difference in group delay.

\section{Conclusion} 

We have proposed a feasible photonic crystal waveguide for realizing superluminal edge states.
A systematic method to produce EPs 
in non-Hermitian photonic crystal waveguides with glide and time reversal symmetry
has been presented,
and the exact EP condition about the permittivity adjustment
is analytically derived by $\bm{k} \cdot \bm{p}$ Hamiltonian.
We have shown that the EPs can be generated by introducing the loss-bias to the system,
and the EPs can be restored by appropriately adjusting the permittivity.
The anomalous dispersion near the EPs leads to the high group velocity of the edge states,
which allows the realization of superluminal propagation.
In addition, the high group velocity of the edge states near the EPs
enlarges the propagation length of the waveguide, even in the lossy and short lifetime system.
We have confirmed that the group velocity contrast of the edge state is increased and corresponding propagation length is enlarged
by appropriately adjusting the permittivity to restore EP.
Our scheme for generating and restoring EPs in photonic crystal waveguides
can be applied to the silicon photonic crystal slabs
by loading graphene and adjusting the position of triangular holes.
We have confirmed that the group velocity $v_g$ is accelerated to $25$ times of that of w/o graphene
near the restored EPs,
thereby realizing the superluminal edge state ($v_g = 3.3c$).
Importantly, the propagation length is also elongated by the same ratio,
directly resulting from the fast light effect.

Our study pave the way to realize the observation of superluminal propagation states 
in photonic crystal waveguide.
The feasibility and integration capabilities, 
combined with graphene's reversibility, make it a promising candidate for optical devices.
In addition, the fast-light on-tip nanodevices have the potential for the development of
control of light-matter interaction as well as the traditional slow-light waveguide,
opening up new possibilities for the application of optical computing.
Furthermore, in addition to the anomalous dispersion in the edge states,
our structure has additional intriguing aspect about the topological properties. 
The proposed structure is based on a photonic crystal waveguide, 
known for exhibiting edge states with valley-polarized characteristics 
and robustness against bending \cite{Shalaev2019, Yoshimi:21}. 
Our structure involves uniform permittivity perturbations 
on both sides of a glide plane, 
which preserves the symmetry of the effective Hamiltonian for the bulk. 
This results in maintaining a finite Berry curvature near the K and K' points
in the band structure. 
Consequently, the edge states near the EPs retain valley-polarized topological features. 
Although a detailed discussion of valley topology is beyond the scope of this work, 
our proposed structure holds potential for application in valley photonic devices,
and provides a pathway for developing innovative photonic devices.

%%%%%%%%%%%%%%%%%%%%%%%%%%%%%%%%%%

\begin{acknowledgments}
This work was supported by the Japan Society for the Promotion of Science (Grant number JP20H05641, JP21K14551, 24K01377, 24H02232 and 24H00400).
\end{acknowledgments}

\appendix

\section{The design of heterostructure in other lattice or hole shapes}
\label{sec:appendix_design_heterostructure}

%%%%% Figure S1 %%%%%
\begin{figure*}
  \includegraphics[width=17cm]{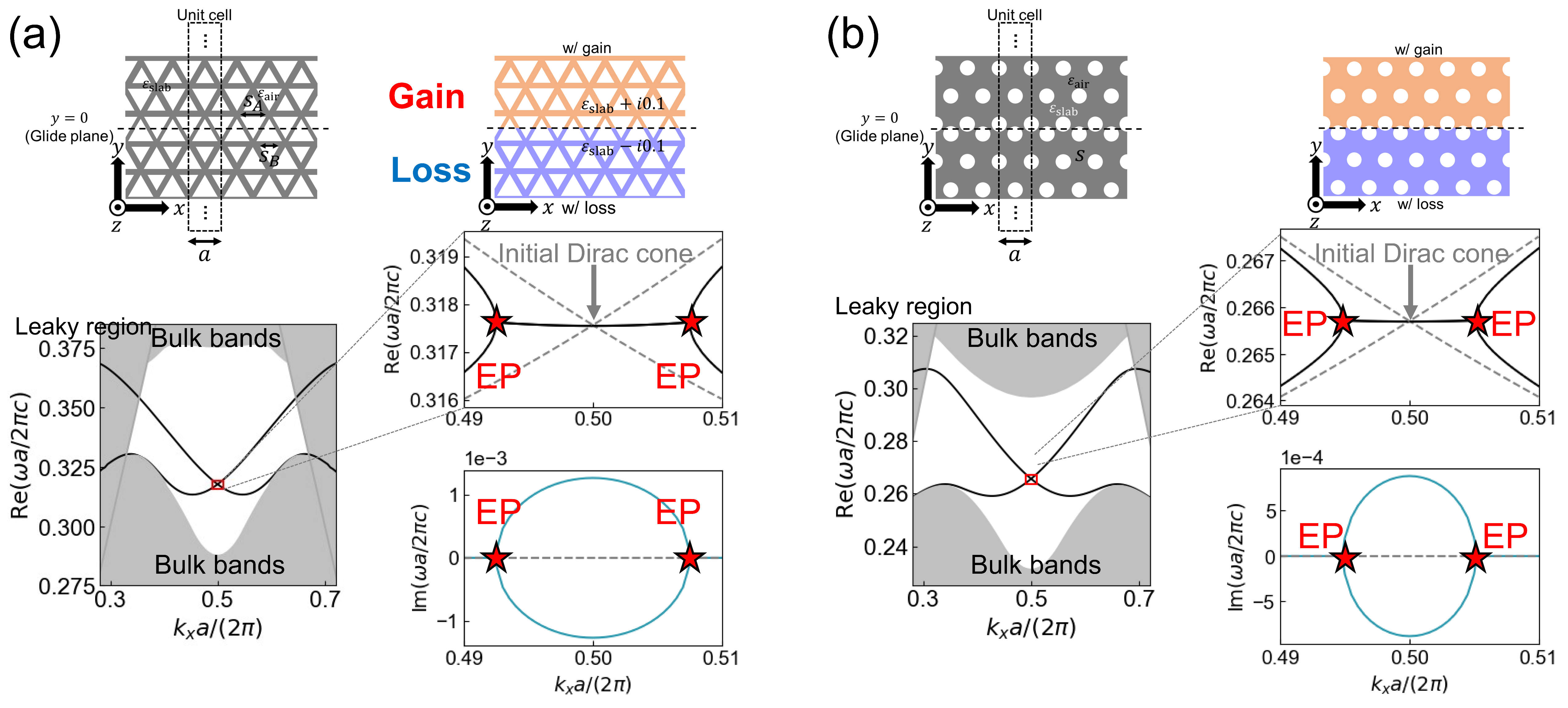}
  \caption{\label{fig_2Dotheredge}
      Interface geometry and band dispersion curves of the photonic crystal hetero-structure
      with (a) the honeycomb lattice and triangular holes, and 
      (b) the triangular lattice and circular holes.
      }
\end{figure*}
Our scheme for generating EPs from a symmetry oriented Dirac point 
can be applied to other lattice structures and hole shapes 
as long as the spatial photonic crystal geometry has the glide symmetry. 
Figure \ref{fig_2Dotheredge} (a) shows the schematic and the band dispersion 
in the honeycomb lattice photonic crystal. 
The slab permittivity and the amount of the perturbation
are the same as in the triangular lattice. 
It is obvious that the EPs generated by the non-Hermitian perturbation 
on both sides of the glide plane are the same as in the triangular lattice. 
On the other hand, the generated EPs are outside the band gap 
when we adopt the honeycomb lattice, 
and it is difficult to bring the EPs inside the band gap by some kind of tuning. 
Figure \ref{fig_2Dotheredge} (b) shows the results when we adopt the triangular lattice with circular hole, 
and there is no significant difference compared with the use of triangular holes. 
Thus, our proposed method for generating symmetry-oriented EPs 
can be applied to photonic crystal with different lattice structures. 
The structure of the triangular lattice and triangular holes shown in Fig. \ref{fig_2Dwg} 
will be analyzed in the following section
because it is suitable for fast-light waveguide applications 
since the EP is inside the band gap, the band slope (group velocity) near the Dirac point is large, 
and the band gap is wide.

\section{$H_z$ distribution of the eigenmodes}
\label{sec:appendix_Hz_distribution}

%%%%% Figure S2 %%%%%
\begin{figure}
  \includegraphics[width=8cm]{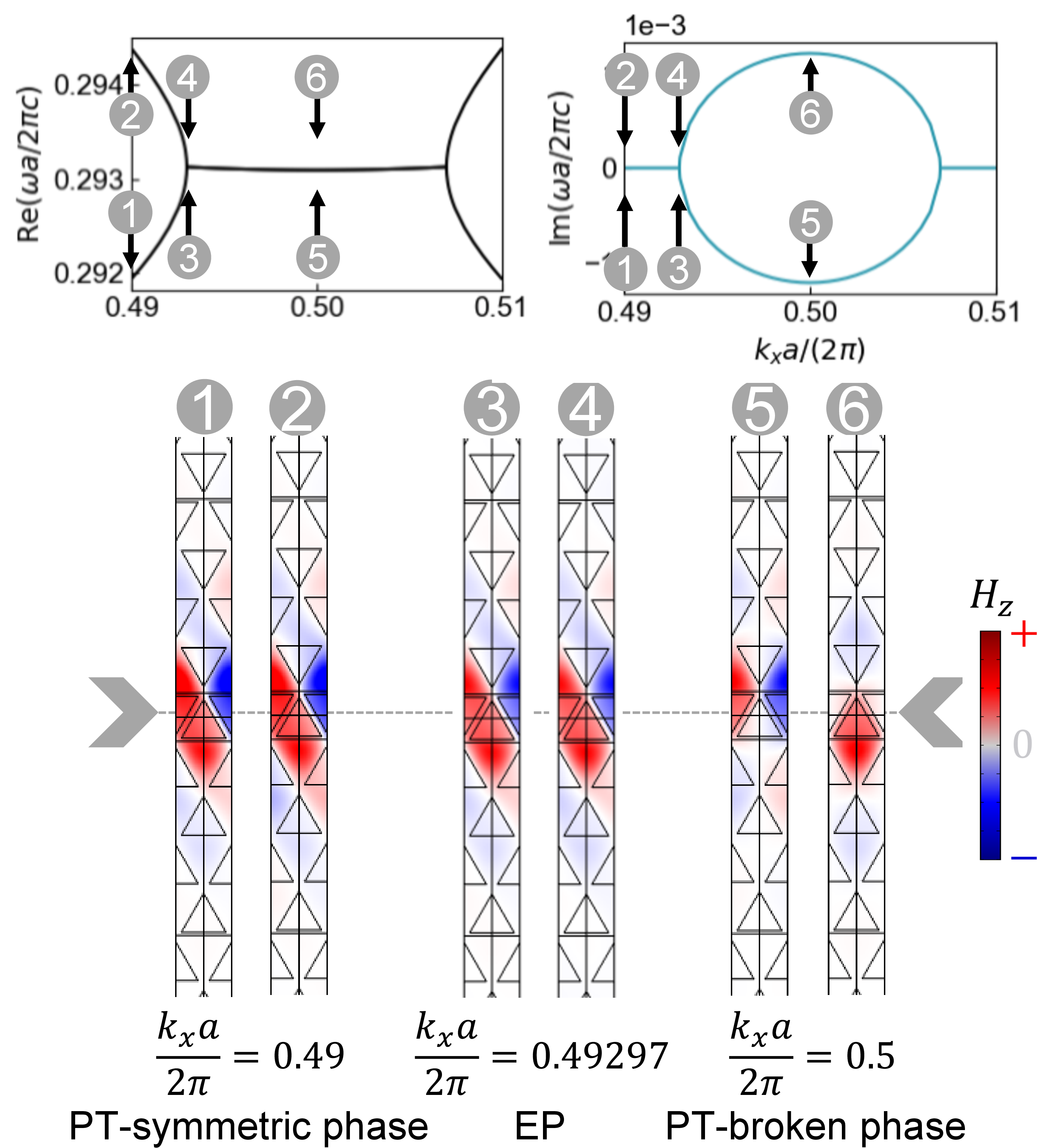}
  \caption{\label{fig_2DHz}
      $H_z$ distribution of the eigenmodes in the photonic crystal heterostructure
      at (a) $k_x a / (2\pi) = 0.49$, (b) $0.49297$ (EP), and (c) $0.5$.
  }
\end{figure}
In the context of a two-band system with PT symmetry, 
the phase transition of eigenvalues and eigenstates through the EP is 
known as the PT phase transition \cite{Feng2017}. 
Since we now focus on the TE mode, 
the two eigenmodes are characterized by their $H_z$ distributions.
For three points, namely outside the EP ($k_x = 0.98 \times 2\pi / a$), 
at the EP ($k_x = 0.49297 \times 2\pi / a$), 
and at the original Dirac point ($k_x = \pi / a$), 
we present the $H_z$ distributions of the two bands at each $k_x$ point in Fig. \ref{fig_2DHz}. 
We discuss the presence or absence of PT symmetry for each case. 
At $k_x = 0.49 \times 2\pi / a$, the $H_z$ distribution is symmetric about $y=0$, 
which is consistent with both eigenvalues being real. 
In addition, the symmetry under simultaneous space and time reversal operations, 
i.e., $(\hat{G}\hat{T})H_z^{(i)} = H_z^{(i)}$ and $i=1,2$ denotes the label of eigenmodes, 
is evident from the figure (PT-symmetric phase).
Similar features are observed in the EP ($k_x = 0.49297 \times 2\pi / a$), 
but the two eigenmodes show complete coalescence, 
which distinguishes it from the $k_x = 0.49 \times 2\pi / a$ case. 
%(it is difficult to see the difference in the behavior of modes A and B from the static image. )
At $k_x = \pi / a$, the $H_z$ distribution under PT operations is no longer symmetric, 
but shows an asymmetric distribution with one side favoring gain and the other side favoring loss
(PT-broken phase).
In summary, the $H_z$ distributions of the eigenmodes in the heterostructure shown in Fig. \ref{fig_2Dwg} 
satisfy the properties of a general PT-symmetric system.

\section{Derivation of the $\bm{k} \cdot \bm{p}$ Hamiltonian}
\label{sec:appendix_kp_hamiltonian}
Here, we derive the $\bm{k} \cdot \bm{p}$ Hamiltonian from the Maxwell's equations.
The details of the derivation are shown in Ref. \cite{PhysRevLett.116.203902}.

The Maxwell's equations in the photonic crystal are given by \cite{sakoda2005optical}
\begin{align}
  \label{eq:maxwell}
  \bm{\nabla} \times \bm{\nabla} \times \bm{E}_{nk}(\bm{r}) 
  = \left[ \varepsilon(\bm{r}) + \Delta \varepsilon(\bm{r}) \right]
  {\left( \frac{\omega_{nk}}{c} \right)}^2 \bm{E}_{nk}(\bm{r}) 
\end{align}
Here, $\varepsilon(\mathbf{r})$ is the permittivity of initial photonic crystal,
$\Delta\varepsilon(\mathbf{r}) := \Delta\varepsilon_{\mathbf{r}} + i\Delta\varepsilon_{\mathbf{i}}(\mathbf{r})$ is 
the perturbation of the permittivity.
$\omega_{nk}$ and $\bm{E}_{nk}(\bm{r})$ are the eigenfrequency and electric filed profile of the photonic crystal, 
respectively.
And, $\omega_{n}^{(0)}$ is the unperturbed eigenfrequency.
Let $U_n^{(0)}$ be the unperturbed eigenenergy, 
and the bases $\bm{E}_{nk}^{(0)}(\bm{r})$ holds the orthonormality condition:
\begin{align}
  \frac{\varepsilon_0}{4}
  \int_{\mathrm{u.c.}} 
  \varepsilon(\bm{r})
  \bm{E}_{nk}^{(0)*}(\bm{r}) \cdot \bm{E}_{mk}^{(0)}(\bm{r}) \mathrm{d} \bm{r} 
  = U_n^{(0)} \delta_{nm}
\end{align}
Then, we define the normalized eigenmodes $\bm{\psi}_{nk}^{(0)}(\bm{r})$ as
\begin{align}
  \bm{\psi}_{nk}^{(0)} (\bm{r}) 
  = \frac{ \sqrt{\varepsilon_0} }{ 2\sqrt{U_n^{(0)}} } \bm{E}_{nk}^{(0)}(\bm{r})
\end{align}
and we get the following equation:
\begin{align}
  \label{eq:maxwell_psi}
  \bm{\nabla} \times \bm{\nabla} \times \bm{\psi}_{nk}(\bm{r}) 
 = \left[ \varepsilon(\bm{r}) + \Delta \varepsilon(\bm{r}) \right]
 {\left( \frac{\omega_{nk}}{c} \right)}^2 \bm{\psi}_{nk}(\bm{r}) 
 \\
 \int_{\mathrm{u.c.}} 
 \varepsilon(\bm{r})
 \bm{\psi}_{nk}^{(0)*}(\bm{r}) \cdot \bm{\psi}_{mk}^{(0)}(\bm{r}) \mathrm{d} \bm{r} 
 = \delta_{nm}
\end{align}
Like the electric field, 
the magnetic field is also normalised as 
\begin{align}
  \label{eq:def_varphi}
  \bm{\varphi}_{nk}^{(0)} (\bm{r}) 
  = \frac{ \sqrt{\mu_0} }{ 2\sqrt{U_n^{(0)}} } \bm{H}_{nk}^{(0)}(\bm{r})
\end{align}

Next, we expand the wave function $\bm{\psi}_{nk}(\bm{r})$  
when $\Delta \varepsilon(\bm{r})$ is finite by the wave function 
at $\bm{k}={[k_0,0,0]}^\top, \ k_0 = \pi/a$ 
with $\Delta \varepsilon(\bm{r}) = 0$:
\begin{align}
  \label{eq:expansion_psi}
  \bm{\psi}_{nk}(\bm{r})
  &=
  \sum_j C_{nj}(k) e^{i \bm{\delta k}\cdot \bm{r}} \bm{\psi}_{jk_0}^{(0)}(\bm{r})
\end{align}
Here, $\bm{\delta k} = \bm{k} - {[k_0,0,0]}^\top$ and $C_{nj}(k)$ is the expansion coefficient.
Substitute Eq. \eqref{eq:expansion_psi} into Eq. \eqref{eq:maxwell_psi} and we get
\begin{align}
  &\sum_j C_{nj}(k) e^{i \bm{\delta k}\cdot \bm{r}}
  A \bm{\psi}_{jk_0}^{(0)}(\bm{r})
  \nonumber\\
  &= 
  {\left( \frac{\omega_{nk}}{c} \right)}^2 
  \left[ \varepsilon(\bm{r}) + \Delta \varepsilon(\bm{r}) \right]
  \sum_j C_{nj}(k) e^{i \bm{\delta k}\cdot \bm{r}} \bm{\psi}_{jk_0}^{(0)}(\bm{r})
  \nonumber\\
  A &:= -
  \bm{\delta k} \times \bm{\delta k} \times 
  + 
  i \left(
    \bm{\nabla} \times \bm{\delta k} \times 
    \right. 
  \nonumber\\
    &\left. \qquad \qquad \qquad 
    + 
    \bm{\delta k} \times \bm{\nabla} \times 
  \right) + 
  \bm{\nabla} \times \bm{\nabla} \times 
\end{align}
Furthermore, by multiplying from the left 
by $e^{-i \bm{\delta k}\cdot \bm{r}} \bm{\psi}_{ik_0}^{(0)*}(\bm{r})$
and integrating over the unit cell,
the eigenvalue equation determining the band structure is obtained.

\begin{align}
  &\sum_j 
  \left[
    {\left( \frac{\omega_{jk}^{(0)}}{c} \right)}^2 \delta_{ij} +
    P_{ij}' + Q_{ij}' 
  \right]
  C_{nj}(k) \nonumber\\
  &\qquad \qquad \qquad \qquad
  = {\left( \frac{\omega_{nk}}{c} \right)}^2
  \sum_j
  \left(
    \delta_{ij} + G_{ij}
  \right)
\end{align}
Here, 
$\displaystyle \bm{\nabla} \times \bm{\nabla} \times \bm{\psi}_{jk}^{(0)}(\bm{r}) 
= \varepsilon(\bm{r}) {\left( \frac{\omega_{nk}^{(0)}}{c} \right)}^2 \bm{\psi}_{jk}^{(0)}(\bm{r})$
is used in the derivation.
$P_{ij}'$, $G_{ij}$, $Q'_{ij}$ are defined as follows:
\begin{align}
  P_{ij}'
  &= \int_{\mathrm{u.c.}} 
  \left(
    \omega_j^{(0)} \bm{\psi}_{ik_0}^{(0)*}(\bm{r}) \times \bm{\varphi}_{jk_0}^{(0)}(\bm{r})
    \right. 
    \nonumber\\
    &\qquad \qquad \qquad
    \left.
    - \omega_i^{(0)} \bm{\varphi}_{ik_0}^{(0)*}(\bm{r}) \times \bm{\psi}_{jk_0}^{(0)}(\bm{r})    
  \right) 
  \cdot \bm{\delta k} \,
  \mathrm{d}^3r \nonumber\\
  &=:
  \bm{P}_{ij} \cdot \bm{\delta k}
  \\
  G_{ij} &=
  \int_{\mathrm{u.c.}} 
  \Delta \varepsilon (\bm{r})
  \bm{\psi}_{ik_0}^{(0)*}(\bm{r}) \cdot \bm{\psi}_{jk_0}^{(0)}(\bm{r})
  \mathrm{d}^3r
  \\
  Q'_{ij} &=
  -\int_{\mathrm{u.c.}} 
  \bm{\psi}_{ik_0}^{(0)*}(\bm{r}) \cdot \left( \bm{\delta k} \times \bm{\delta k} \times \bm{\psi}_{jk_0}^{(0)}(\bm{r}) \right)
  \mathrm{d}^3r \nonumber\\
  &=
  {(\delta k_x)}^2
  \int_{\mathrm{u.c.}} 
  \left(
  {\left( \psi_{ik_0}^{(0)*}(\bm{r}) \right)}_y
  {\left( \psi_{jk_0}^{(0)}(\bm{r}) \right)}_y
  \right. \nonumber\\
  &\qquad \qquad \qquad \qquad \left. 
  + {\left( \psi_{ik_0}^{(0)*}(\bm{r}) \right)}_z
  {\left( \psi_{jk_0}^{(0)}(\bm{r}) \right)}_z
  \right)
  \mathrm{d}^3r \nonumber\\
  &=: {(\delta k_x)}^2 Q_{ij}
\end{align}

\section{The EP condition with the hole shift in $\bm{k} \cdot \bm{p}$ Hamiltonian}
\label{sec:appendix_EP_hole_shift}

%%%%% Figure S3 %%%%%
\begin{figure}
  \includegraphics[width=8cm]{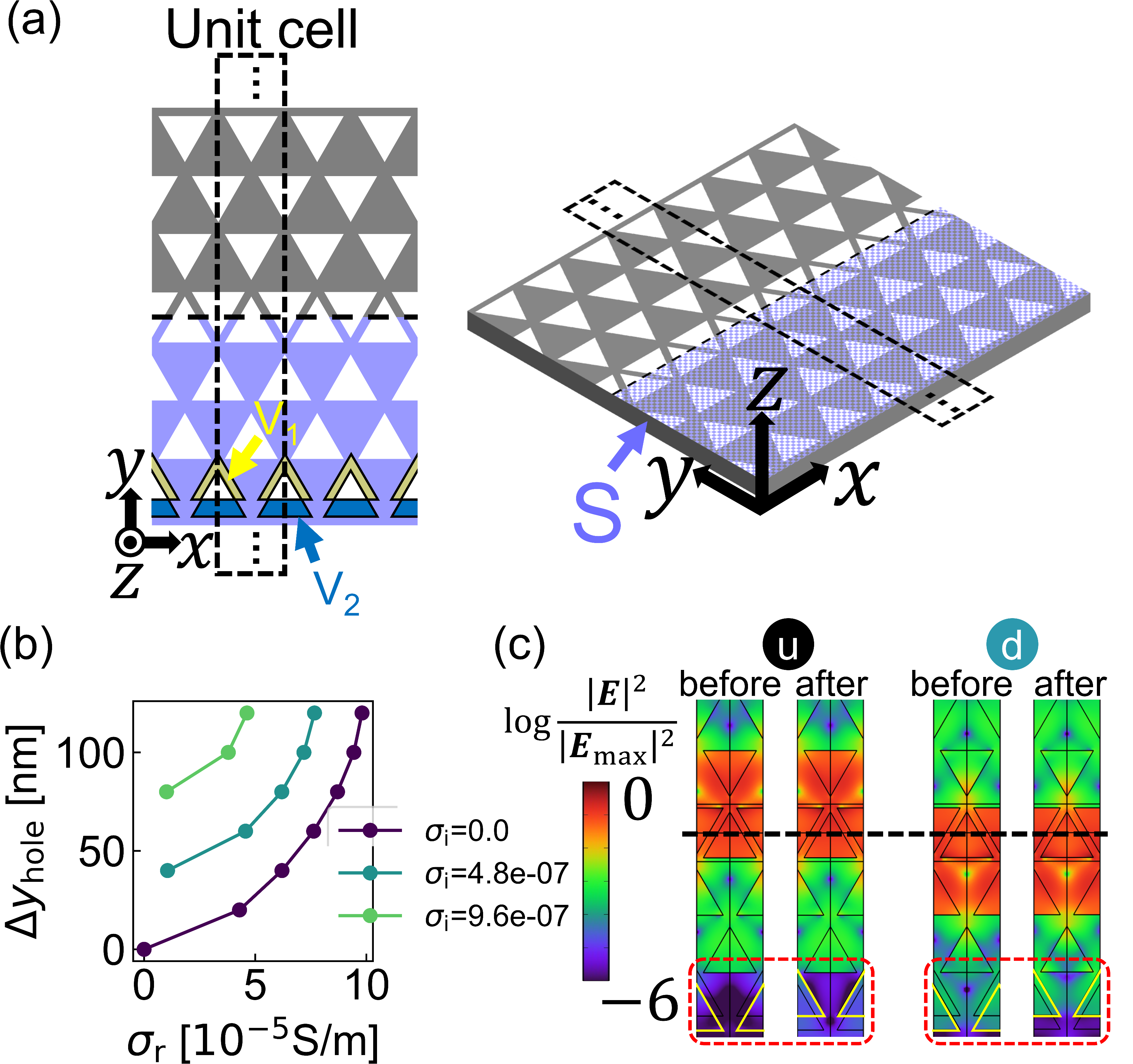}
  \caption{\label{fig_EP_hole_shift}
  (a) The definition of the domain $V_1, V_2$ and the surface $S$. 
  (b) The required triangular hole shift $\Delta y_\mathrm{hole}$ to restore the EPs
  as a function of a real part of conductivity $\sigma_r$ at each imaginary part of conductivity $\sigma_i$ [S/m].
  (c) The electric field distribution of the eigenmodes at $k_x = \pi/a$ before and after the hole shift.
  }
\end{figure}

We have shown in Section \ref{sec:restoration_EP} of that the EP can be restored by appropriately adjusting the real part of the permittivity, 
and adopted the shift of the triangular holes in the realistic structure as a method instead of the permittivity adjustment in Section \ref{sec:graphene}.
Here, we theoretically investigate the effect of the triangular hole shift and show that the triangular hole shift offset the effect of the permittivity adjustment.

The following describes the setup for treating the triangular hole shift and graphene loading as perturbations in the permittivity. A domain $V_1, V_2$ and surface $S$ are defined as shown in Fig. \ref{fig_EP_hole_shift}(a). The domain $V_1$ and $V_2$ are defined as the regions where the permittivity changes from silicon to air and from air to silicon, respectively, when the triangular holes are shifted by $\Delta y_\mathrm{hole}$. 
Moreover, the surface $S$ is defined as the region where the graphene is loaded.
The elements of the matrix $G$ defined by the Eq.\eqref{eq:def_matG} are the sum of the contributions from the domains $V_1, V_2$ and the surface $S$ as follows:

\begin{align}
  G_\mathrm{\alpha \beta} &= G_\mathrm{\alpha \beta}^{(V_1)} + G_\mathrm{\alpha \beta}^{(V_2)} + G_\mathrm{\alpha \beta}^{(S)}, \quad \alpha, \beta = \mathrm{d, u}
\end{align}
Here, $G_\mathrm{ud}, G_\mathrm{du}$ are zero because all the domains and surfaces have mirror symmetry with respect to the $y$-axis of the unit cell.
The contributions from the domains $V_1, V_2$ are given by
\begin{align}
  G_\mathrm{\alpha \beta}^{(V_1)} &= (\varepsilon_\mathrm{air}-\varepsilon_\mathrm{slab}) \int_{V_1} \bm{\psi}_{\alpha k_0}^{(0)*}(\bm{r}) \cdot \bm{\psi}_{\beta k_0}^{(0)}(\bm{r}) \mathrm{d}^3 r \\
  G_\mathrm{\alpha \beta}^{(V_2)} &= (\varepsilon_\mathrm{slab} - \varepsilon_\mathrm{air}) \int_{V_2} \bm{\psi}_{\alpha k_0}^{(0)*}(\bm{r}) \cdot \bm{\psi}_{\beta k_0}^{(0)}(\bm{r}) \mathrm{d}^3 r
\end{align}
The permittivity of the silicon and air are denoted as $\varepsilon_\mathrm{slab}$ and $\varepsilon_\mathrm{air}$, respectively.
To simplify the notation, we introduce the following quantities:
\begin{align}
  I_\mathrm{\alpha \beta}^{(V)} &= \int_{V} \bm{\psi}_{\alpha k_0}^{(0)*}(\bm{r}) \cdot \bm{\psi}_{\beta k_0}^{(0)}(\bm{r}) \mathrm{d}^3 r
\end{align}
Then, the contributions from $V_1$ and $V_2$ are collectively denoted as $Y_\mathrm{d}, Y_\mathrm{u}$:
\begin{align}
  \label{eq:def_Yd}
  Y_\mathrm{d} := G_\mathrm{dd}^{(V_1)} + G_\mathrm{dd}^{(V_2)} 
  &= (\varepsilon_\mathrm{slab} - \varepsilon_\mathrm{air}) (I_\mathrm{dd}^{(V_2)}-I_\mathrm{dd}^{(V_1)}) \\
  \label{eq:def_Yu}
  Y_\mathrm{u} := G_\mathrm{uu}^{(V_1)} + G_\mathrm{uu}^{(V_2)}
  &= (\varepsilon_\mathrm{slab} - \varepsilon_\mathrm{air}) (I_\mathrm{uu}^{(V_2)}-I_\mathrm{uu}^{(V_1)})
\end{align}
Next, we consider the contribution from the surface $S$.
Since the complex conductivity of the graphene is given by $\sigma = \sigma_r + i \sigma_i$, the permittivity perturbation due to the graphene loading is given by 
\begin{align}
  \Delta \varepsilon (\bm{r}) = -i \frac{\sigma}{t\omega \varepsilon_0}, \quad \bm{r} \in S
\end{align}
Here, $t$ is the thickness of the graphene, $\omega$ is the angular frequency, and $\varepsilon_0$ is the vacuum permittivity. For simplicity, we will ignore wavelength dispersion of the frequency.
Then, the contribution from the surface $S$ is given by
\begin{align}
  G_\mathrm{\alpha \beta}^{(S)} &= \int_{S} \frac{-i\sigma}{t \omega \varepsilon_0} \bm{\psi}_{\alpha k_0}^{(0)*}(\bm{r}) \cdot \bm{\psi}_{\beta k_0}^{(0)}(\bm{r})  t \, \mathrm{d}^2 r \nonumber\\
  &= -\frac{i\sigma}{\omega \varepsilon_0} \int_{S} \bm{\psi}_{\alpha k_0}^{(0)*}(\bm{r}) \cdot \bm{\psi}_{\beta k_0}^{(0)}(\bm{r})  \, \mathrm{d}^2 r \nonumber\\
  &=: -i\sigma L_{\alpha \beta}
\end{align}
where
\begin{align}
  \label{eq:def_matL}
  L_{\alpha \beta} = \frac{1}{\omega \varepsilon_0} \int_{S} \bm{\psi}_{\alpha k_0}^{(0)*}(\bm{r}) \cdot \bm{\psi}_{\beta k_0}^{(0)}(\bm{r})  \, \mathrm{d}^2 r
\end{align}
The contributions from the domains $V_1, V_2$ and the surface $S$ are collectively denoted as
\begin{align}
  G_\mathrm{dd} &= Y_\mathrm{d} + (\sigma_i - i\sigma_r) L_{\mathrm{dd}}, \\
  G_\mathrm{uu} &= Y_\mathrm{u} + (\sigma_i - i\sigma_r) L_{\mathrm{uu}}
\end{align}
All $V_1$, $V_2$ and $S$ depend on the hole shift $\Delta y_\mathrm{hole}$, but the value of $L_{\alpha \beta}$ is almost constant since the contribution of the hole shift in the graphene-loaded region is small.

Next, we calculate the components to evaluate the EP condition in Eq. \eqref{eq:ep_condition_matG}.
\begin{align}
  4\Delta G &= G_\mathrm{dd} - G_\mathrm{uu} \nonumber\\
  &= Y_\mathrm{d} - Y_\mathrm{u} + (\sigma_i - i\sigma_r) (L_{\mathrm{dd}} - L_{\mathrm{uu}}) \nonumber\\
  &=: \Delta y + (\sigma_i - i\sigma_r) \Delta L
\end{align}
\begin{align}
  &(1+G_\mathrm{dd})(1+G_\mathrm{uu}) \nonumber\\
  &= (1+Y_\mathrm{d} + \sigma_i L_{\mathrm{dd}} - i \sigma_r L_{\mathrm{dd}})(1+Y_\mathrm{u} + \sigma_i L_{\mathrm{uu}} - i \sigma_r L_{\mathrm{uu}}) \nonumber\\
  &= (1+Y_\mathrm{d} + \sigma_i L_{\mathrm{dd}} )(1+Y_\mathrm{u} + \sigma_i L_{\mathrm{uu}}) - \sigma_r^2 L_{\mathrm{dd}} L_{\mathrm{uu}} \nonumber\\ 
  &\quad -i \sigma_r \left(
    (1+Y_\mathrm{d} + \sigma_i L_{\mathrm{dd}}) L_{\mathrm{uu}} + (1+Y_\mathrm{u} + \sigma_i L_{\mathrm{uu}}) L_{\mathrm{dd}}
  \right)
  \nonumber\\
\end{align}
By substituting the above equations into Eq. \eqref{eq:ep_condition_matG}, we obtain the EP condition as follows:
\begin{align}
  \label{eq:ep_condition_hole_shift}
  \sigma_r^2 &= \frac{
    m^2 l - 2m \Delta L (1+Y_\mathrm{d} + \sigma_i L_{\mathrm{dd}} )(1+Y_\mathrm{u} + \sigma_i L_{\mathrm{uu}})
  }{
    l{(\Delta L)}^2 - 2m \Delta L L_{\mathrm{dd}} L_{\mathrm{uu}}
  } 
\end{align}
where
\begin{align}
  l &= (1+Y_\mathrm{d} + \sigma_i L_{\mathrm{dd}}) L_{\mathrm{uu}} + (1+Y_\mathrm{u} + \sigma_i L_{\mathrm{uu}}) L_{\mathrm{dd}} \\
  m &= Y_\mathrm{d} - Y_\mathrm{u} + \sigma_i \Delta L
\end{align}
Eq. \eqref{eq:ep_condition_hole_shift} assures that the EP can be restored by shifting the triangular holes.
From the expression for $Y_\mathrm{d} + \sigma_i L_{\mathrm{dd}}$ and $Y_\mathrm{u} + \sigma_i L_{\mathrm{uu}}$ that appears in Eq.\eqref{eq:ep_condition_hole_shift}, it can be seen that the triangular hole shift $\Delta y_\mathrm{hole}$ and the imaginary part of the graphene conductivity (i.e., the real part of the permittivity) 
have the opposite effect on the permittivity perturbation.
Generally, $I_\mathrm{dd}^{(V_2)} < I_\mathrm{dd}^{(V_1)}$, $I_\mathrm{uu}^{(V_2)} < I_\mathrm{uu}^{(V_1)}$ hold since the edge states are localized near the glide plane. Consequently, $Y_\mathrm{d}, Y_\mathrm{u}$ defined by Eqs. \eqref{eq:def_Yd}, \eqref{eq:def_Yu} are negative.  
Therefore, the triangular hole shift $\Delta y_\mathrm{hole}$ has the same effect as a net negative tuning of the real part of the permittivity.
Furthermore, due to the difference in the electric field distribution between modes $\mathrm{u}$ and $\mathrm{d}$, the permittivity perturbation due to the hole shift affects only the $\mathrm{d}$ mode.
Note that since graphene has a positive imaginary part of its electrical conductivity, i.e., a positive real part of its permittivity, it is not possible to restore the EP only using graphene.

Figure \ref{fig_EP_hole_shift} (b) shows the relationship between the real part of the conductivity and the triangular hole shift required to satisfy the EP condition when the imaginary part of the conductivity is fixed. When the imaginary part of the conductivity is zero, the triangular hole shift that satisfies the EP condition is approximately proportional to the square of the real part of the conductivity. As previously mentioned, a positive imaginary part of the conductivity has an opposite effect to the triangular hole shift, so the larger the imaginary part of the conductivity, the narrower the range of the real part of the conductivity that can be canceled out by the triangular hole shift. In practice, the imaginary part of the conductivity for graphene is about $\sigma_i = 2 \times 10^{-6}$ [S/m] \cite{10.1063/1.2891452} so within the range of perturbation theory, it is difficult to restore EP using only the hole shift. In actual FEM calculations, however, the EP restoration has been confirmed with a hole shift of about $\Delta y_\mathrm{hole} = 105$ nm, but this is believed to be due to the large permittivity contrast $\varepsilon_\mathrm{slab} - \varepsilon_\mathrm{air} \approx 11$, which causes changes in the electric field distribution, making the contribution of the actual triangular hole shift larger than the linear perturbation effect. The change in the electric field distribution due to the triangular hole shift is confirmed in Fig. \ref{fig_EP_hole_shift} (c).

\section{Restoration of EPs and group velocity in graphene-Loaded structures with misalignment}
\label{sec:appendix_graphene_misalignment}

%%%%% Figure S4 %%%%%
\begin{figure}
  \includegraphics[width=8cm]{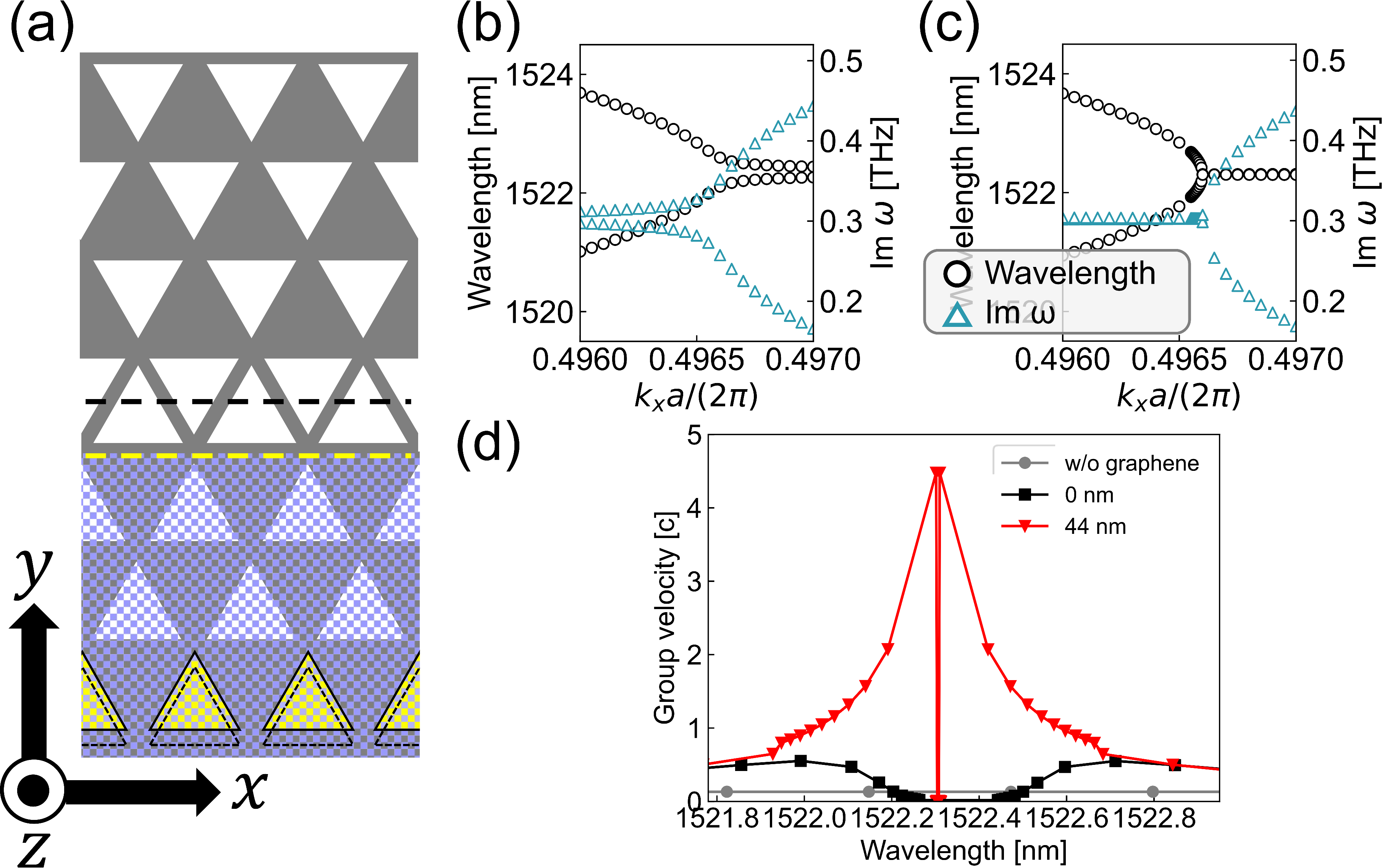}
  \caption{\label{fig_graphene_misalignment}
  (a) Schematic of the photonic crystal slab loaded with 5 sheet of graphene, 
  where the graphene is misaligned by $\sqrt{3}a/4$ below the glide plane (black dashed line).
  (b), (c) The band dispersion of the photonic crystal slab with the graphene misalignment, 
  (b) without no hole shift and (c) with the hole shift $\Delta y_\mathrm{hole} = 44$ nm.
  (d) The group velocity of the edge states near the EPs with and without the hole shift.
  }
\end{figure}

The graphene-loaded structure proposed in Section \label{sec:graphene} assumes that graphene is placed along the glide-symmetry plane. From the perspective of fabrication and measurement, it is important to consider the effects on EP and group velocity when the graphene alignment is offset. In practice, even if the graphene position is misaligned, our scheme for EP restoration can still be applied without issue. Figure \ref{fig_graphene_misalignment} (a) shows the case where the graphene loading position is shifted by $\sqrt{3}a/4$ below the glide plane. For this structure, the band diagram and group velocity for both the case without triangular hole shifts and with a shift of 44 nm are shown in Fig. \ref{fig_graphene_misalignment} (b) and (c), respectively. As is evident from the figures, even when the graphene position is misaligned, the EP is restored, and the singularity in group velocity is recovered as shown in Fig. \ref{fig_graphene_misalignment} (d). Mathematically, the misalignment of the graphene loading position is equivalent to changing the magnitude of graphene's complex permittivity. In other words, it ultimately affects the value of the integral $L_{\alpha \beta}$ in Eq. \ref{eq:def_matL}. Therefore, by adjusting the value of the triangular hole shift or permittivity shift required to cancel the EP, it is possible to restore the EP in exactly the same manner as in the case without misalignment.

\bibliography{reference}% Produces the bibliography via BibTeX.

\providecommand{\noopsort}[1]{}\providecommand{\singleletter}[1]{#1}%
\begin{thebibliography}{49}%
\makeatletter
\providecommand \@ifxundefined [1]{%
 \@ifx{#1\undefined}
}%
\providecommand \@ifnum [1]{%
 \ifnum #1\expandafter \@firstoftwo
 \else \expandafter \@secondoftwo
 \fi
}%
\providecommand \@ifx [1]{%
 \ifx #1\expandafter \@firstoftwo
 \else \expandafter \@secondoftwo
 \fi
}%
\providecommand \natexlab [1]{#1}%
\providecommand \enquote  [1]{``#1''}%
\providecommand \bibnamefont  [1]{#1}%
\providecommand \bibfnamefont [1]{#1}%
\providecommand \citenamefont [1]{#1}%
\providecommand \href@noop [0]{\@secondoftwo}%
\providecommand \href [0]{\begingroup \@sanitize@url \@href}%
\providecommand \@href[1]{\@@startlink{#1}\@@href}%
\providecommand \@@href[1]{\endgroup#1\@@endlink}%
\providecommand \@sanitize@url [0]{\catcode `\\12\catcode `\$12\catcode `\&12\catcode `\#12\catcode `\^12\catcode `\_12\catcode `\%12\relax}%
\providecommand \@@startlink[1]{}%
\providecommand \@@endlink[0]{}%
\providecommand \url  [0]{\begingroup\@sanitize@url \@url }%
\providecommand \@url [1]{\endgroup\@href {#1}{\urlprefix }}%
\providecommand \urlprefix  [0]{URL }%
\providecommand \Eprint [0]{\href }%
\providecommand \doibase [0]{https://doi.org/}%
\providecommand \selectlanguage [0]{\@gobble}%
\providecommand \bibinfo  [0]{\@secondoftwo}%
\providecommand \bibfield  [0]{\@secondoftwo}%
\providecommand \translation [1]{[#1]}%
\providecommand \BibitemOpen [0]{}%
\providecommand \bibitemStop [0]{}%
\providecommand \bibitemNoStop [0]{.\EOS\space}%
\providecommand \EOS [0]{\spacefactor3000\relax}%
\providecommand \BibitemShut  [1]{\csname bibitem#1\endcsname}%
\let\auto@bib@innerbib\@empty
%</preamble>
\bibitem [{\citenamefont {Bender}\ and\ \citenamefont {Boettcher}(1998)}]{PhysRevLett.80.5243}%
  \BibitemOpen
  \bibfield  {author} {\bibinfo {author} {\bibfnamefont {C.~M.}\ \bibnamefont {Bender}}\ and\ \bibinfo {author} {\bibfnamefont {S.}~\bibnamefont {Boettcher}},\ }\bibfield  {title} {\bibinfo {title} {Real spectra in non-hermitian hamiltonians having $\mathcal{P}\mathcal{T}$ symmetry},\ }\href {https://doi.org/10.1103/PhysRevLett.80.5243} {\bibfield  {journal} {\bibinfo  {journal} {Phys. Rev. Lett.}\ }\textbf {\bibinfo {volume} {80}},\ \bibinfo {pages} {5243} (\bibinfo {year} {1998})}\BibitemShut {NoStop}%
\bibitem [{\citenamefont {Bender}(2007)}]{Bender_2007}%
  \BibitemOpen
  \bibfield  {author} {\bibinfo {author} {\bibfnamefont {C.~M.}\ \bibnamefont {Bender}},\ }\bibfield  {title} {\bibinfo {title} {Making sense of non-hermitian hamiltonians},\ }\href {https://doi.org/10.1088/0034-4885/70/6/R03} {\bibfield  {journal} {\bibinfo  {journal} {Reports on Progress in Physics}\ }\textbf {\bibinfo {volume} {70}},\ \bibinfo {pages} {947} (\bibinfo {year} {2007})}\BibitemShut {NoStop}%
\bibitem [{\citenamefont {Parto}\ \emph {et~al.}(2018)\citenamefont {Parto}, \citenamefont {Wittek}, \citenamefont {Hodaei}, \citenamefont {Harari}, \citenamefont {Bandres}, \citenamefont {Ren}, \citenamefont {Rechtsman}, \citenamefont {Segev}, \citenamefont {Christodoulides},\ and\ \citenamefont {Khajavikhan}}]{PhysRevLett.120.113901}%
  \BibitemOpen
  \bibfield  {author} {\bibinfo {author} {\bibfnamefont {M.}~\bibnamefont {Parto}}, \bibinfo {author} {\bibfnamefont {S.}~\bibnamefont {Wittek}}, \bibinfo {author} {\bibfnamefont {H.}~\bibnamefont {Hodaei}}, \bibinfo {author} {\bibfnamefont {G.}~\bibnamefont {Harari}}, \bibinfo {author} {\bibfnamefont {M.~A.}\ \bibnamefont {Bandres}}, \bibinfo {author} {\bibfnamefont {J.}~\bibnamefont {Ren}}, \bibinfo {author} {\bibfnamefont {M.~C.}\ \bibnamefont {Rechtsman}}, \bibinfo {author} {\bibfnamefont {M.}~\bibnamefont {Segev}}, \bibinfo {author} {\bibfnamefont {D.~N.}\ \bibnamefont {Christodoulides}},\ and\ \bibinfo {author} {\bibfnamefont {M.}~\bibnamefont {Khajavikhan}},\ }\bibfield  {title} {\bibinfo {title} {Edge-mode lasing in 1d topological active arrays},\ }\href {https://doi.org/10.1103/PhysRevLett.120.113901} {\bibfield  {journal} {\bibinfo  {journal} {Phys. Rev. Lett.}\ }\textbf {\bibinfo {volume} {120}},\ \bibinfo {pages} {113901} (\bibinfo {year} {2018})}\BibitemShut {NoStop}%
\bibitem [{\citenamefont {Takata}\ \emph {et~al.}(2021)\citenamefont {Takata}, \citenamefont {Nozaki}, \citenamefont {Kuramochi}, \citenamefont {Matsuo}, \citenamefont {Takeda}, \citenamefont {Fujii}, \citenamefont {Kita}, \citenamefont {Shinya},\ and\ \citenamefont {Notomi}}]{Takata:21}%
  \BibitemOpen
  \bibfield  {author} {\bibinfo {author} {\bibfnamefont {K.}~\bibnamefont {Takata}}, \bibinfo {author} {\bibfnamefont {K.}~\bibnamefont {Nozaki}}, \bibinfo {author} {\bibfnamefont {E.}~\bibnamefont {Kuramochi}}, \bibinfo {author} {\bibfnamefont {S.}~\bibnamefont {Matsuo}}, \bibinfo {author} {\bibfnamefont {K.}~\bibnamefont {Takeda}}, \bibinfo {author} {\bibfnamefont {T.}~\bibnamefont {Fujii}}, \bibinfo {author} {\bibfnamefont {S.}~\bibnamefont {Kita}}, \bibinfo {author} {\bibfnamefont {A.}~\bibnamefont {Shinya}},\ and\ \bibinfo {author} {\bibfnamefont {M.}~\bibnamefont {Notomi}},\ }\bibfield  {title} {\bibinfo {title} {Observing exceptional point degeneracy of radiation with electrically pumped photonic crystal coupled-nanocavity lasers},\ }\href {https://doi.org/10.1364/OPTICA.412596} {\bibfield  {journal} {\bibinfo  {journal} {Optica}\ }\textbf {\bibinfo {volume} {8}},\ \bibinfo {pages} {184} (\bibinfo {year} {2021})}\BibitemShut {NoStop}%
\bibitem [{\citenamefont {Feng}\ \emph {et~al.}(2017)\citenamefont {Feng}, \citenamefont {El-Ganainy},\ and\ \citenamefont {Ge}}]{Feng2017}%
  \BibitemOpen
  \bibfield  {author} {\bibinfo {author} {\bibfnamefont {L.}~\bibnamefont {Feng}}, \bibinfo {author} {\bibfnamefont {R.}~\bibnamefont {El-Ganainy}},\ and\ \bibinfo {author} {\bibfnamefont {L.}~\bibnamefont {Ge}},\ }\bibfield  {title} {\bibinfo {title} {Non-hermitian photonics based on parity--time symmetry},\ }\href {https://doi.org/10.1038/s41566-017-0031-1} {\bibfield  {journal} {\bibinfo  {journal} {Nature Photonics}\ }\textbf {\bibinfo {volume} {11}},\ \bibinfo {pages} {752} (\bibinfo {year} {2017})}\BibitemShut {NoStop}%
\bibitem [{\citenamefont {Regensburger}\ \emph {et~al.}(2012)\citenamefont {Regensburger}, \citenamefont {Bersch}, \citenamefont {Miri}, \citenamefont {Onishchukov}, \citenamefont {Christodoulides},\ and\ \citenamefont {Peschel}}]{Regensburger2012}%
  \BibitemOpen
  \bibfield  {author} {\bibinfo {author} {\bibfnamefont {A.}~\bibnamefont {Regensburger}}, \bibinfo {author} {\bibfnamefont {C.}~\bibnamefont {Bersch}}, \bibinfo {author} {\bibfnamefont {M.-A.}\ \bibnamefont {Miri}}, \bibinfo {author} {\bibfnamefont {G.}~\bibnamefont {Onishchukov}}, \bibinfo {author} {\bibfnamefont {D.~N.}\ \bibnamefont {Christodoulides}},\ and\ \bibinfo {author} {\bibfnamefont {U.}~\bibnamefont {Peschel}},\ }\bibfield  {title} {\bibinfo {title} {Parity-time synthetic photonic lattices},\ }\href {https://doi.org/10.1038/nature11298} {\bibfield  {journal} {\bibinfo  {journal} {Nature}\ }\textbf {\bibinfo {volume} {488}},\ \bibinfo {pages} {167} (\bibinfo {year} {2012})}\BibitemShut {NoStop}%
\bibitem [{\citenamefont {{\"O}zdemir}\ \emph {et~al.}(2019)\citenamefont {{\"O}zdemir}, \citenamefont {Rotter}, \citenamefont {Nori},\ and\ \citenamefont {Yang}}]{Ozdemir2019}%
  \BibitemOpen
  \bibfield  {author} {\bibinfo {author} {\bibfnamefont {{\c{S}}.~K.}\ \bibnamefont {{\"O}zdemir}}, \bibinfo {author} {\bibfnamefont {S.}~\bibnamefont {Rotter}}, \bibinfo {author} {\bibfnamefont {F.}~\bibnamefont {Nori}},\ and\ \bibinfo {author} {\bibfnamefont {L.}~\bibnamefont {Yang}},\ }\bibfield  {title} {\bibinfo {title} {Parity--time symmetry and exceptional points in photonics},\ }\href {https://doi.org/10.1038/s41563-019-0304-9} {\bibfield  {journal} {\bibinfo  {journal} {Nature Materials}\ }\textbf {\bibinfo {volume} {18}},\ \bibinfo {pages} {783} (\bibinfo {year} {2019})}\BibitemShut {NoStop}%
\bibitem [{\citenamefont {Peng}\ \emph {et~al.}(2014)\citenamefont {Peng}, \citenamefont {{\"O}zdemir}, \citenamefont {Lei}, \citenamefont {Monifi}, \citenamefont {Gianfreda}, \citenamefont {Long}, \citenamefont {Fan}, \citenamefont {Nori}, \citenamefont {Bender},\ and\ \citenamefont {Yang}}]{Peng2014}%
  \BibitemOpen
  \bibfield  {author} {\bibinfo {author} {\bibfnamefont {B.}~\bibnamefont {Peng}}, \bibinfo {author} {\bibfnamefont {{\c{S}}.~K.}\ \bibnamefont {{\"O}zdemir}}, \bibinfo {author} {\bibfnamefont {F.}~\bibnamefont {Lei}}, \bibinfo {author} {\bibfnamefont {F.}~\bibnamefont {Monifi}}, \bibinfo {author} {\bibfnamefont {M.}~\bibnamefont {Gianfreda}}, \bibinfo {author} {\bibfnamefont {G.~L.}\ \bibnamefont {Long}}, \bibinfo {author} {\bibfnamefont {S.}~\bibnamefont {Fan}}, \bibinfo {author} {\bibfnamefont {F.}~\bibnamefont {Nori}}, \bibinfo {author} {\bibfnamefont {C.~M.}\ \bibnamefont {Bender}},\ and\ \bibinfo {author} {\bibfnamefont {L.}~\bibnamefont {Yang}},\ }\bibfield  {title} {\bibinfo {title} {Parity--time-symmetric whispering-gallery microcavities},\ }\href {https://doi.org/10.1038/nphys2927} {\bibfield  {journal} {\bibinfo  {journal} {Nature Physics}\ }\textbf {\bibinfo {volume} {10}},\ \bibinfo {pages} {394} (\bibinfo {year} {2014})}\BibitemShut {NoStop}%
\bibitem [{\citenamefont {Feng}\ \emph {et~al.}(2014)\citenamefont {Feng}, \citenamefont {Wong}, \citenamefont {Ma}, \citenamefont {Wang},\ and\ \citenamefont {Zhang}}]{doi:10.1126/science.1258479}%
  \BibitemOpen
  \bibfield  {author} {\bibinfo {author} {\bibfnamefont {L.}~\bibnamefont {Feng}}, \bibinfo {author} {\bibfnamefont {Z.~J.}\ \bibnamefont {Wong}}, \bibinfo {author} {\bibfnamefont {R.-M.}\ \bibnamefont {Ma}}, \bibinfo {author} {\bibfnamefont {Y.}~\bibnamefont {Wang}},\ and\ \bibinfo {author} {\bibfnamefont {X.}~\bibnamefont {Zhang}},\ }\bibfield  {title} {\bibinfo {title} {Single-mode laser by parity-time symmetry breaking},\ }\href {https://doi.org/10.1126/science.1258479} {\bibfield  {journal} {\bibinfo  {journal} {Science}\ }\textbf {\bibinfo {volume} {346}},\ \bibinfo {pages} {972} (\bibinfo {year} {2014})}\BibitemShut {NoStop}%
\bibitem [{\citenamefont {Hodaei}\ \emph {et~al.}(2014)\citenamefont {Hodaei}, \citenamefont {Miri}, \citenamefont {Heinrich}, \citenamefont {Christodoulides},\ and\ \citenamefont {Khajavikhan}}]{doi:10.1126/science.1258480}%
  \BibitemOpen
  \bibfield  {author} {\bibinfo {author} {\bibfnamefont {H.}~\bibnamefont {Hodaei}}, \bibinfo {author} {\bibfnamefont {M.-A.}\ \bibnamefont {Miri}}, \bibinfo {author} {\bibfnamefont {M.}~\bibnamefont {Heinrich}}, \bibinfo {author} {\bibfnamefont {D.~N.}\ \bibnamefont {Christodoulides}},\ and\ \bibinfo {author} {\bibfnamefont {M.}~\bibnamefont {Khajavikhan}},\ }\bibfield  {title} {\bibinfo {title} {Parity-time-symmetric microring lasers},\ }\href {https://doi.org/10.1126/science.1258480} {\bibfield  {journal} {\bibinfo  {journal} {Science}\ }\textbf {\bibinfo {volume} {346}},\ \bibinfo {pages} {975} (\bibinfo {year} {2014})}\BibitemShut {NoStop}%
\bibitem [{\citenamefont {Guo}\ \emph {et~al.}(2009)\citenamefont {Guo}, \citenamefont {Salamo}, \citenamefont {Duchesne}, \citenamefont {Morandotti}, \citenamefont {Volatier-Ravat}, \citenamefont {Aimez}, \citenamefont {Siviloglou},\ and\ \citenamefont {Christodoulides}}]{PhysRevLett.103.093902}%
  \BibitemOpen
  \bibfield  {author} {\bibinfo {author} {\bibfnamefont {A.}~\bibnamefont {Guo}}, \bibinfo {author} {\bibfnamefont {G.~J.}\ \bibnamefont {Salamo}}, \bibinfo {author} {\bibfnamefont {D.}~\bibnamefont {Duchesne}}, \bibinfo {author} {\bibfnamefont {R.}~\bibnamefont {Morandotti}}, \bibinfo {author} {\bibfnamefont {M.}~\bibnamefont {Volatier-Ravat}}, \bibinfo {author} {\bibfnamefont {V.}~\bibnamefont {Aimez}}, \bibinfo {author} {\bibfnamefont {G.~A.}\ \bibnamefont {Siviloglou}},\ and\ \bibinfo {author} {\bibfnamefont {D.~N.}\ \bibnamefont {Christodoulides}},\ }\bibfield  {title} {\bibinfo {title} {Observation of $\mathcal{P}\mathcal{T}$-symmetry breaking in complex optical potentials},\ }\href {https://doi.org/10.1103/PhysRevLett.103.093902} {\bibfield  {journal} {\bibinfo  {journal} {Phys. Rev. Lett.}\ }\textbf {\bibinfo {volume} {103}},\ \bibinfo {pages} {093902} (\bibinfo {year} {2009})}\BibitemShut {NoStop}%
\bibitem [{\citenamefont {Szameit}\ \emph {et~al.}(2011)\citenamefont {Szameit}, \citenamefont {Rechtsman}, \citenamefont {Bahat-Treidel},\ and\ \citenamefont {Segev}}]{PhysRevA.84.021806}%
  \BibitemOpen
  \bibfield  {author} {\bibinfo {author} {\bibfnamefont {A.}~\bibnamefont {Szameit}}, \bibinfo {author} {\bibfnamefont {M.~C.}\ \bibnamefont {Rechtsman}}, \bibinfo {author} {\bibfnamefont {O.}~\bibnamefont {Bahat-Treidel}},\ and\ \bibinfo {author} {\bibfnamefont {M.}~\bibnamefont {Segev}},\ }\bibfield  {title} {\bibinfo {title} {$\mathcal{P}\mathcal{T}$-symmetry in honeycomb photonic lattices},\ }\href {https://doi.org/10.1103/PhysRevA.84.021806} {\bibfield  {journal} {\bibinfo  {journal} {Phys. Rev. A}\ }\textbf {\bibinfo {volume} {84}},\ \bibinfo {pages} {021806} (\bibinfo {year} {2011})}\BibitemShut {NoStop}%
\bibitem [{\citenamefont {Baba}(2008)}]{Baba2008}%
  \BibitemOpen
  \bibfield  {author} {\bibinfo {author} {\bibfnamefont {T.}~\bibnamefont {Baba}},\ }\bibfield  {title} {\bibinfo {title} {Slow light in photonic crystals},\ }\href {https://doi.org/10.1038/nphoton.2008.146} {\bibfield  {journal} {\bibinfo  {journal} {Nature Photonics}\ }\textbf {\bibinfo {volume} {2}},\ \bibinfo {pages} {465} (\bibinfo {year} {2008})}\BibitemShut {NoStop}%
\bibitem [{\citenamefont {Yariv}\ \emph {et~al.}(1999)\citenamefont {Yariv}, \citenamefont {Xu}, \citenamefont {Lee},\ and\ \citenamefont {Scherer}}]{Yariv:99}%
  \BibitemOpen
  \bibfield  {author} {\bibinfo {author} {\bibfnamefont {A.}~\bibnamefont {Yariv}}, \bibinfo {author} {\bibfnamefont {Y.}~\bibnamefont {Xu}}, \bibinfo {author} {\bibfnamefont {R.~K.}\ \bibnamefont {Lee}},\ and\ \bibinfo {author} {\bibfnamefont {A.}~\bibnamefont {Scherer}},\ }\bibfield  {title} {\bibinfo {title} {Coupled-resonator optical waveguide:?a proposal and analysis},\ }\href {https://doi.org/10.1364/OL.24.000711} {\bibfield  {journal} {\bibinfo  {journal} {Opt. Lett.}\ }\textbf {\bibinfo {volume} {24}},\ \bibinfo {pages} {711} (\bibinfo {year} {1999})}\BibitemShut {NoStop}%
\bibitem [{\citenamefont {Olivier}\ \emph {et~al.}(2001)\citenamefont {Olivier}, \citenamefont {Smith}, \citenamefont {Rattier}, \citenamefont {Benisty}, \citenamefont {Weisbuch}, \citenamefont {Krauss}, \citenamefont {Houdr\'{e}},\ and\ \citenamefont {Oesterl\'{e}}}]{Olivier:01}%
  \BibitemOpen
  \bibfield  {author} {\bibinfo {author} {\bibfnamefont {S.}~\bibnamefont {Olivier}}, \bibinfo {author} {\bibfnamefont {C.}~\bibnamefont {Smith}}, \bibinfo {author} {\bibfnamefont {M.}~\bibnamefont {Rattier}}, \bibinfo {author} {\bibfnamefont {H.}~\bibnamefont {Benisty}}, \bibinfo {author} {\bibfnamefont {C.}~\bibnamefont {Weisbuch}}, \bibinfo {author} {\bibfnamefont {T.}~\bibnamefont {Krauss}}, \bibinfo {author} {\bibfnamefont {R.}~\bibnamefont {Houdr\'{e}}},\ and\ \bibinfo {author} {\bibfnamefont {U.}~\bibnamefont {Oesterl\'{e}}},\ }\bibfield  {title} {\bibinfo {title} {Miniband transmission in a photonic crystal coupled-resonator optical waveguide},\ }\href {https://doi.org/10.1364/OL.26.001019} {\bibfield  {journal} {\bibinfo  {journal} {Opt. Lett.}\ }\textbf {\bibinfo {volume} {26}},\ \bibinfo {pages} {1019} (\bibinfo {year} {2001})}\BibitemShut {NoStop}%
\bibitem [{\citenamefont {Notomi}\ \emph {et~al.}(2008)\citenamefont {Notomi}, \citenamefont {Kuramochi},\ and\ \citenamefont {Tanabe}}]{Notomi2008}%
  \BibitemOpen
  \bibfield  {author} {\bibinfo {author} {\bibfnamefont {M.}~\bibnamefont {Notomi}}, \bibinfo {author} {\bibfnamefont {E.}~\bibnamefont {Kuramochi}},\ and\ \bibinfo {author} {\bibfnamefont {T.}~\bibnamefont {Tanabe}},\ }\bibfield  {title} {\bibinfo {title} {Large-scale arrays of ultrahigh-q coupled nanocavities},\ }\href {https://doi.org/10.1038/nphoton.2008.226} {\bibfield  {journal} {\bibinfo  {journal} {Nature Photonics}\ }\textbf {\bibinfo {volume} {2}},\ \bibinfo {pages} {741} (\bibinfo {year} {2008})}\BibitemShut {NoStop}%
\bibitem [{\citenamefont {Vlasov}\ \emph {et~al.}(2005)\citenamefont {Vlasov}, \citenamefont {O'Boyle}, \citenamefont {Hamann},\ and\ \citenamefont {McNab}}]{Vlasov2005}%
  \BibitemOpen
  \bibfield  {author} {\bibinfo {author} {\bibfnamefont {Y.~A.}\ \bibnamefont {Vlasov}}, \bibinfo {author} {\bibfnamefont {M.}~\bibnamefont {O'Boyle}}, \bibinfo {author} {\bibfnamefont {H.~F.}\ \bibnamefont {Hamann}},\ and\ \bibinfo {author} {\bibfnamefont {S.~J.}\ \bibnamefont {McNab}},\ }\bibfield  {title} {\bibinfo {title} {Active control of slow light on a chip with photonic crystal waveguides},\ }\href {https://doi.org/10.1038/nature04210} {\bibfield  {journal} {\bibinfo  {journal} {Nature}\ }\textbf {\bibinfo {volume} {438}},\ \bibinfo {pages} {65} (\bibinfo {year} {2005})}\BibitemShut {NoStop}%
\bibitem [{\citenamefont {Yoshimi}\ \emph {et~al.}(2020)\citenamefont {Yoshimi}, \citenamefont {Yamaguchi}, \citenamefont {Ota}, \citenamefont {Arakawa},\ and\ \citenamefont {Iwamoto}}]{Yoshimi:20}%
  \BibitemOpen
  \bibfield  {author} {\bibinfo {author} {\bibfnamefont {H.}~\bibnamefont {Yoshimi}}, \bibinfo {author} {\bibfnamefont {T.}~\bibnamefont {Yamaguchi}}, \bibinfo {author} {\bibfnamefont {Y.}~\bibnamefont {Ota}}, \bibinfo {author} {\bibfnamefont {Y.}~\bibnamefont {Arakawa}},\ and\ \bibinfo {author} {\bibfnamefont {S.}~\bibnamefont {Iwamoto}},\ }\bibfield  {title} {\bibinfo {title} {Slow light waveguides in topological valley photonic crystals},\ }\href {https://doi.org/10.1364/OL.391764} {\bibfield  {journal} {\bibinfo  {journal} {Opt. Lett.}\ }\textbf {\bibinfo {volume} {45}},\ \bibinfo {pages} {2648} (\bibinfo {year} {2020})}\BibitemShut {NoStop}%
\bibitem [{\citenamefont {Yoshimi}\ \emph {et~al.}(2021)\citenamefont {Yoshimi}, \citenamefont {Yamaguchi}, \citenamefont {Katsumi}, \citenamefont {Ota}, \citenamefont {Arakawa},\ and\ \citenamefont {Iwamoto}}]{Yoshimi:21}%
  \BibitemOpen
  \bibfield  {author} {\bibinfo {author} {\bibfnamefont {H.}~\bibnamefont {Yoshimi}}, \bibinfo {author} {\bibfnamefont {T.}~\bibnamefont {Yamaguchi}}, \bibinfo {author} {\bibfnamefont {R.}~\bibnamefont {Katsumi}}, \bibinfo {author} {\bibfnamefont {Y.}~\bibnamefont {Ota}}, \bibinfo {author} {\bibfnamefont {Y.}~\bibnamefont {Arakawa}},\ and\ \bibinfo {author} {\bibfnamefont {S.}~\bibnamefont {Iwamoto}},\ }\bibfield  {title} {\bibinfo {title} {Experimental demonstration of topological slow light waveguides in valley photonic crystals},\ }\href {https://doi.org/10.1364/OE.422962} {\bibfield  {journal} {\bibinfo  {journal} {Opt. Express}\ }\textbf {\bibinfo {volume} {29}},\ \bibinfo {pages} {13441} (\bibinfo {year} {2021})}\BibitemShut {NoStop}%
\bibitem [{\citenamefont {Brillouin}(1960)}]{Brillouin:1960tos}%
  \BibitemOpen
  \bibfield  {author} {\bibinfo {author} {\bibfnamefont {L.}~\bibnamefont {Brillouin}},\ }\href@noop {} {\emph {\bibinfo {title} {{Wave propagation and group velocity}}}},\ Vol.~\bibinfo {volume} {8}\ (\bibinfo  {publisher} {Academic Press},\ \bibinfo {address} {New York, London},\ \bibinfo {year} {1960})\BibitemShut {NoStop}%
\bibitem [{\citenamefont {Wang}\ \emph {et~al.}(2000)\citenamefont {Wang}, \citenamefont {Kuzmich},\ and\ \citenamefont {Dogariu}}]{Wang2000}%
  \BibitemOpen
  \bibfield  {author} {\bibinfo {author} {\bibfnamefont {L.~J.}\ \bibnamefont {Wang}}, \bibinfo {author} {\bibfnamefont {A.}~\bibnamefont {Kuzmich}},\ and\ \bibinfo {author} {\bibfnamefont {A.}~\bibnamefont {Dogariu}},\ }\bibfield  {title} {\bibinfo {title} {Gain-assisted superluminal light propagation},\ }\href {https://doi.org/10.1038/35018520} {\bibfield  {journal} {\bibinfo  {journal} {Nature}\ }\textbf {\bibinfo {volume} {406}},\ \bibinfo {pages} {277} (\bibinfo {year} {2000})}\BibitemShut {NoStop}%
\bibitem [{\citenamefont {Bigelow}\ \emph {et~al.}(2003)\citenamefont {Bigelow}, \citenamefont {Lepeshkin},\ and\ \citenamefont {Boyd}}]{doi:10.1126/science.1084429}%
  \BibitemOpen
  \bibfield  {author} {\bibinfo {author} {\bibfnamefont {M.~S.}\ \bibnamefont {Bigelow}}, \bibinfo {author} {\bibfnamefont {N.~N.}\ \bibnamefont {Lepeshkin}},\ and\ \bibinfo {author} {\bibfnamefont {R.~W.}\ \bibnamefont {Boyd}},\ }\bibfield  {title} {\bibinfo {title} {Superluminal and slow light propagation in a room-temperature solid},\ }\href {https://doi.org/10.1126/science.1084429} {\bibfield  {journal} {\bibinfo  {journal} {Science}\ }\textbf {\bibinfo {volume} {301}},\ \bibinfo {pages} {200} (\bibinfo {year} {2003})}\BibitemShut {NoStop}%
\bibitem [{\citenamefont {Ziolkowski}(2001)}]{PhysRevE.63.046604}%
  \BibitemOpen
  \bibfield  {author} {\bibinfo {author} {\bibfnamefont {R.~W.}\ \bibnamefont {Ziolkowski}},\ }\bibfield  {title} {\bibinfo {title} {Superluminal transmission of information through an electromagnetic metamaterial},\ }\href {https://doi.org/10.1103/PhysRevE.63.046604} {\bibfield  {journal} {\bibinfo  {journal} {Phys. Rev. E}\ }\textbf {\bibinfo {volume} {63}},\ \bibinfo {pages} {046604} (\bibinfo {year} {2001})}\BibitemShut {NoStop}%
\bibitem [{\citenamefont {Xu}\ \emph {et~al.}(2012)\citenamefont {Xu}, \citenamefont {Cheng}, \citenamefont {Xi}, \citenamefont {Zhang}, \citenamefont {Moser}, \citenamefont {Shen}, \citenamefont {Xu}, \citenamefont {Huang}, \citenamefont {Zhang}, \citenamefont {Yu}, \citenamefont {Zhang},\ and\ \citenamefont {Chen}}]{PhysRevLett.109.223903}%
  \BibitemOpen
  \bibfield  {author} {\bibinfo {author} {\bibfnamefont {S.}~\bibnamefont {Xu}}, \bibinfo {author} {\bibfnamefont {X.}~\bibnamefont {Cheng}}, \bibinfo {author} {\bibfnamefont {S.}~\bibnamefont {Xi}}, \bibinfo {author} {\bibfnamefont {R.}~\bibnamefont {Zhang}}, \bibinfo {author} {\bibfnamefont {H.~O.}\ \bibnamefont {Moser}}, \bibinfo {author} {\bibfnamefont {Z.}~\bibnamefont {Shen}}, \bibinfo {author} {\bibfnamefont {Y.}~\bibnamefont {Xu}}, \bibinfo {author} {\bibfnamefont {Z.}~\bibnamefont {Huang}}, \bibinfo {author} {\bibfnamefont {X.}~\bibnamefont {Zhang}}, \bibinfo {author} {\bibfnamefont {F.}~\bibnamefont {Yu}}, \bibinfo {author} {\bibfnamefont {B.}~\bibnamefont {Zhang}},\ and\ \bibinfo {author} {\bibfnamefont {H.}~\bibnamefont {Chen}},\ }\bibfield  {title} {\bibinfo {title} {Experimental demonstration of a free-space cylindrical cloak without superluminal propagation},\ }\href {https://doi.org/10.1103/PhysRevLett.109.223903} {\bibfield  {journal} {\bibinfo  {journal} {Phys. Rev. Lett.}\ }\textbf {\bibinfo {volume} {109}},\ \bibinfo {pages} {223903} (\bibinfo {year} {2012})}\BibitemShut {NoStop}%
\bibitem [{\citenamefont {Longhi}(2013)}]{PhysRevA.88.052102}%
  \BibitemOpen
  \bibfield  {author} {\bibinfo {author} {\bibfnamefont {S.}~\bibnamefont {Longhi}},\ }\bibfield  {title} {\bibinfo {title} {Convective and absolute $\mathcal{PT}$-symmetry breaking in tight-binding lattices},\ }\href {https://doi.org/10.1103/PhysRevA.88.052102} {\bibfield  {journal} {\bibinfo  {journal} {Phys. Rev. A}\ }\textbf {\bibinfo {volume} {88}},\ \bibinfo {pages} {052102} (\bibinfo {year} {2013})}\BibitemShut {NoStop}%
\bibitem [{\citenamefont {Schomerus}\ and\ \citenamefont {Wiersig}(2014)}]{PhysRevA.90.053819}%
  \BibitemOpen
  \bibfield  {author} {\bibinfo {author} {\bibfnamefont {H.}~\bibnamefont {Schomerus}}\ and\ \bibinfo {author} {\bibfnamefont {J.}~\bibnamefont {Wiersig}},\ }\bibfield  {title} {\bibinfo {title} {Non-hermitian-transport effects in coupled-resonator optical waveguides},\ }\href {https://doi.org/10.1103/PhysRevA.90.053819} {\bibfield  {journal} {\bibinfo  {journal} {Phys. Rev. A}\ }\textbf {\bibinfo {volume} {90}},\ \bibinfo {pages} {053819} (\bibinfo {year} {2014})}\BibitemShut {NoStop}%
\bibitem [{\citenamefont {Takata}\ and\ \citenamefont {Notomi}(2017)}]{PhysRevApplied.7.054023}%
  \BibitemOpen
  \bibfield  {author} {\bibinfo {author} {\bibfnamefont {K.}~\bibnamefont {Takata}}\ and\ \bibinfo {author} {\bibfnamefont {M.}~\bibnamefont {Notomi}},\ }\bibfield  {title} {\bibinfo {title} {$\mathcal{P}\mathcal{T}$-symmetric coupled-resonator waveguide based on buried heterostructure nanocavities},\ }\href {https://doi.org/10.1103/PhysRevApplied.7.054023} {\bibfield  {journal} {\bibinfo  {journal} {Phys. Rev. Appl.}\ }\textbf {\bibinfo {volume} {7}},\ \bibinfo {pages} {054023} (\bibinfo {year} {2017})}\BibitemShut {NoStop}%
\bibitem [{\citenamefont {Takata}\ \emph {et~al.}(2022)\citenamefont {Takata}, \citenamefont {Roberts}, \citenamefont {Shinya},\ and\ \citenamefont {Notomi}}]{PhysRevA.105.013523}%
  \BibitemOpen
  \bibfield  {author} {\bibinfo {author} {\bibfnamefont {K.}~\bibnamefont {Takata}}, \bibinfo {author} {\bibfnamefont {N.}~\bibnamefont {Roberts}}, \bibinfo {author} {\bibfnamefont {A.}~\bibnamefont {Shinya}},\ and\ \bibinfo {author} {\bibfnamefont {M.}~\bibnamefont {Notomi}},\ }\bibfield  {title} {\bibinfo {title} {Imaginary couplings in non-hermitian coupled-mode theory: Effects on exceptional points of optical resonators},\ }\href {https://doi.org/10.1103/PhysRevA.105.013523} {\bibfield  {journal} {\bibinfo  {journal} {Phys. Rev. A}\ }\textbf {\bibinfo {volume} {105}},\ \bibinfo {pages} {013523} (\bibinfo {year} {2022})}\BibitemShut {NoStop}%
\bibitem [{\citenamefont {Cui}\ \emph {et~al.}(2019)\citenamefont {Cui}, \citenamefont {Ding}, \citenamefont {Dong},\ and\ \citenamefont {Chan}}]{PhysRevB.100.115412}%
  \BibitemOpen
  \bibfield  {author} {\bibinfo {author} {\bibfnamefont {X.}~\bibnamefont {Cui}}, \bibinfo {author} {\bibfnamefont {K.}~\bibnamefont {Ding}}, \bibinfo {author} {\bibfnamefont {J.-W.}\ \bibnamefont {Dong}},\ and\ \bibinfo {author} {\bibfnamefont {C.~T.}\ \bibnamefont {Chan}},\ }\bibfield  {title} {\bibinfo {title} {Exceptional points and their coalescence of $\mathcal{PT}$-symmetric interface states in photonic crystals},\ }\href {https://doi.org/10.1103/PhysRevB.100.115412} {\bibfield  {journal} {\bibinfo  {journal} {Phys. Rev. B}\ }\textbf {\bibinfo {volume} {100}},\ \bibinfo {pages} {115412} (\bibinfo {year} {2019})}\BibitemShut {NoStop}%
\bibitem [{\citenamefont {Mock}(2020)}]{Mock:20}%
  \BibitemOpen
  \bibfield  {author} {\bibinfo {author} {\bibfnamefont {A.}~\bibnamefont {Mock}},\ }\bibfield  {title} {\bibinfo {title} {Symmetry-engineered waveguide dispersion in $\mathcal{P}\mathcal{T}$ symmetric photonic crystal waveguides},\ }\href {https://doi.org/10.1364/JOSAB.37.000168} {\bibfield  {journal} {\bibinfo  {journal} {J. Opt. Soc. Am. B}\ }\textbf {\bibinfo {volume} {37}},\ \bibinfo {pages} {168} (\bibinfo {year} {2020})}\BibitemShut {NoStop}%
\bibitem [{\citenamefont {Fang}\ \emph {et~al.}(2019)\citenamefont {Fang}, \citenamefont {Ye},\ and\ \citenamefont {Li}}]{Fang_2019}%
  \BibitemOpen
  \bibfield  {author} {\bibinfo {author} {\bibfnamefont {Y.-T.}\ \bibnamefont {Fang}}, \bibinfo {author} {\bibfnamefont {S.-F.}\ \bibnamefont {Ye}},\ and\ \bibinfo {author} {\bibfnamefont {X.-X.}\ \bibnamefont {Li}},\ }\bibfield  {title} {\bibinfo {title} {Unique band coalescence and exceptional points from two-dimensional photonic crystal waveguide},\ }\href {https://doi.org/10.1088/2040-8986/ab0f8d} {\bibfield  {journal} {\bibinfo  {journal} {Journal of Optics}\ }\textbf {\bibinfo {volume} {21}},\ \bibinfo {pages} {055103} (\bibinfo {year} {2019})}\BibitemShut {NoStop}%
\bibitem [{\citenamefont {Iglesias~Mart\'{\i}nez}\ \emph {et~al.}(2022)\citenamefont {Iglesias~Mart\'{\i}nez}, \citenamefont {Laforge}, \citenamefont {Kadic},\ and\ \citenamefont {Laude}}]{PhysRevB.106.064304}%
  \BibitemOpen
  \bibfield  {author} {\bibinfo {author} {\bibfnamefont {J.~A.}\ \bibnamefont {Iglesias~Mart\'{\i}nez}}, \bibinfo {author} {\bibfnamefont {N.}~\bibnamefont {Laforge}}, \bibinfo {author} {\bibfnamefont {M.}~\bibnamefont {Kadic}},\ and\ \bibinfo {author} {\bibfnamefont {V.}~\bibnamefont {Laude}},\ }\bibfield  {title} {\bibinfo {title} {Topological waves guided by a glide-reflection symmetric crystal interface},\ }\href {https://doi.org/10.1103/PhysRevB.106.064304} {\bibfield  {journal} {\bibinfo  {journal} {Phys. Rev. B}\ }\textbf {\bibinfo {volume} {106}},\ \bibinfo {pages} {064304} (\bibinfo {year} {2022})}\BibitemShut {NoStop}%
\bibitem [{\citenamefont {Plotnik}\ \emph {et~al.}(2014)\citenamefont {Plotnik}, \citenamefont {Rechtsman}, \citenamefont {Song}, \citenamefont {Heinrich}, \citenamefont {Zeuner}, \citenamefont {Nolte}, \citenamefont {Lumer}, \citenamefont {Malkova}, \citenamefont {Xu}, \citenamefont {Szameit}, \citenamefont {Chen},\ and\ \citenamefont {Segev}}]{Plotnik2014}%
  \BibitemOpen
  \bibfield  {author} {\bibinfo {author} {\bibfnamefont {Y.}~\bibnamefont {Plotnik}}, \bibinfo {author} {\bibfnamefont {M.~C.}\ \bibnamefont {Rechtsman}}, \bibinfo {author} {\bibfnamefont {D.}~\bibnamefont {Song}}, \bibinfo {author} {\bibfnamefont {M.}~\bibnamefont {Heinrich}}, \bibinfo {author} {\bibfnamefont {J.~M.}\ \bibnamefont {Zeuner}}, \bibinfo {author} {\bibfnamefont {S.}~\bibnamefont {Nolte}}, \bibinfo {author} {\bibfnamefont {Y.}~\bibnamefont {Lumer}}, \bibinfo {author} {\bibfnamefont {N.}~\bibnamefont {Malkova}}, \bibinfo {author} {\bibfnamefont {J.}~\bibnamefont {Xu}}, \bibinfo {author} {\bibfnamefont {A.}~\bibnamefont {Szameit}}, \bibinfo {author} {\bibfnamefont {Z.}~\bibnamefont {Chen}},\ and\ \bibinfo {author} {\bibfnamefont {M.}~\bibnamefont {Segev}},\ }\bibfield  {title} {\bibinfo {title} {Observation of unconventional edge states in `photonic graphene'},\ }\href {https://doi.org/10.1038/nmat3783} {\bibfield  {journal} {\bibinfo  {journal} {Nature Materials}\ }\textbf {\bibinfo {volume} {13}},\ \bibinfo {pages} {57} (\bibinfo {year} {2014})}\BibitemShut {NoStop}%
\bibitem [{\citenamefont {Yoda}\ and\ \citenamefont {Notomi}(2019)}]{yoda2019air}%
  \BibitemOpen
  \bibfield  {author} {\bibinfo {author} {\bibfnamefont {T.}~\bibnamefont {Yoda}}\ and\ \bibinfo {author} {\bibfnamefont {M.}~\bibnamefont {Notomi}},\ }\bibfield  {title} {\bibinfo {title} {Air-hole-type valley photonic crystal slab with simple triangular lattice for valley-contrasting physics},\ }in\ \href@noop {} {\emph {\bibinfo {booktitle} {CLEO: Science and Innovations}}}\ (\bibinfo {organization} {Optica Publishing Group},\ \bibinfo {year} {2019})\ pp.\ \bibinfo {pages} {JTh2A--10}\BibitemShut {NoStop}%
\bibitem [{\citenamefont {S{\"o}llner}\ \emph {et~al.}(2015)\citenamefont {S{\"o}llner}, \citenamefont {Mahmoodian}, \citenamefont {Hansen}, \citenamefont {Midolo}, \citenamefont {Javadi}, \citenamefont {Kir{\v{s}}ansk{\.{e}}}, \citenamefont {Pregnolato}, \citenamefont {El-Ella}, \citenamefont {Lee}, \citenamefont {Song}, \citenamefont {Stobbe},\ and\ \citenamefont {Lodahl}}]{S^^c3^^b6llner2015}%
  \BibitemOpen
  \bibfield  {author} {\bibinfo {author} {\bibfnamefont {I.}~\bibnamefont {S{\"o}llner}}, \bibinfo {author} {\bibfnamefont {S.}~\bibnamefont {Mahmoodian}}, \bibinfo {author} {\bibfnamefont {S.~L.}\ \bibnamefont {Hansen}}, \bibinfo {author} {\bibfnamefont {L.}~\bibnamefont {Midolo}}, \bibinfo {author} {\bibfnamefont {A.}~\bibnamefont {Javadi}}, \bibinfo {author} {\bibfnamefont {G.}~\bibnamefont {Kir{\v{s}}ansk{\.{e}}}}, \bibinfo {author} {\bibfnamefont {T.}~\bibnamefont {Pregnolato}}, \bibinfo {author} {\bibfnamefont {H.}~\bibnamefont {El-Ella}}, \bibinfo {author} {\bibfnamefont {E.~H.}\ \bibnamefont {Lee}}, \bibinfo {author} {\bibfnamefont {J.~D.}\ \bibnamefont {Song}}, \bibinfo {author} {\bibfnamefont {S.}~\bibnamefont {Stobbe}},\ and\ \bibinfo {author} {\bibfnamefont {P.}~\bibnamefont {Lodahl}},\ }\bibfield  {title} {\bibinfo {title} {Deterministic photon--emitter coupling in chiral photonic circuits},\ }\href {https://doi.org/10.1038/nnano.2015.159} {\bibfield  {journal} {\bibinfo  {journal} {Nature Nanotechnology}\ }\textbf {\bibinfo {volume} {10}},\ \bibinfo {pages} {775} (\bibinfo {year} {2015})}\BibitemShut {NoStop}%
\bibitem [{\citenamefont {Mahmoodian}\ \emph {et~al.}(2017)\citenamefont {Mahmoodian}, \citenamefont {Prindal-Nielsen}, \citenamefont {S\"{o}llner}, \citenamefont {Stobbe},\ and\ \citenamefont {Lodahl}}]{Mahmoodian:17}%
  \BibitemOpen
  \bibfield  {author} {\bibinfo {author} {\bibfnamefont {S.}~\bibnamefont {Mahmoodian}}, \bibinfo {author} {\bibfnamefont {K.}~\bibnamefont {Prindal-Nielsen}}, \bibinfo {author} {\bibfnamefont {I.}~\bibnamefont {S\"{o}llner}}, \bibinfo {author} {\bibfnamefont {S.}~\bibnamefont {Stobbe}},\ and\ \bibinfo {author} {\bibfnamefont {P.}~\bibnamefont {Lodahl}},\ }\bibfield  {title} {\bibinfo {title} {Engineering chiral light\&\#x02013;matter interaction in photonic crystal waveguides with slow light},\ }\href {https://doi.org/10.1364/OME.7.000043} {\bibfield  {journal} {\bibinfo  {journal} {Opt. Mater. Express}\ }\textbf {\bibinfo {volume} {7}},\ \bibinfo {pages} {43} (\bibinfo {year} {2017})}\BibitemShut {NoStop}%
\bibitem [{\citenamefont {Liu}\ \emph {et~al.}(2016)\citenamefont {Liu}, \citenamefont {Zhang}, \citenamefont {\"Ozdemir}, \citenamefont {Peng}, \citenamefont {Jing}, \citenamefont {L\"u}, \citenamefont {Li}, \citenamefont {Yang}, \citenamefont {Nori},\ and\ \citenamefont {Liu}}]{PhysRevLett.117.110802}%
  \BibitemOpen
  \bibfield  {author} {\bibinfo {author} {\bibfnamefont {Z.-P.}\ \bibnamefont {Liu}}, \bibinfo {author} {\bibfnamefont {J.}~\bibnamefont {Zhang}}, \bibinfo {author} {\bibfnamefont {i.~m. c.~K.}\ \bibnamefont {\"Ozdemir}}, \bibinfo {author} {\bibfnamefont {B.}~\bibnamefont {Peng}}, \bibinfo {author} {\bibfnamefont {H.}~\bibnamefont {Jing}}, \bibinfo {author} {\bibfnamefont {X.-Y.}\ \bibnamefont {L\"u}}, \bibinfo {author} {\bibfnamefont {C.-W.}\ \bibnamefont {Li}}, \bibinfo {author} {\bibfnamefont {L.}~\bibnamefont {Yang}}, \bibinfo {author} {\bibfnamefont {F.}~\bibnamefont {Nori}},\ and\ \bibinfo {author} {\bibfnamefont {Y.-x.}\ \bibnamefont {Liu}},\ }\bibfield  {title} {\bibinfo {title} {Metrology with $\mathcal{PT}$-symmetric cavities: Enhanced sensitivity near the $\mathcal{PT}$-phase transition},\ }\href {https://doi.org/10.1103/PhysRevLett.117.110802} {\bibfield  {journal} {\bibinfo  {journal} {Phys. Rev. Lett.}\ }\textbf {\bibinfo {volume} {117}},\ \bibinfo {pages} {110802} (\bibinfo {year} {2016})}\BibitemShut {NoStop}%
\bibitem [{\citenamefont {Cerjan}\ \emph {et~al.}(2016)\citenamefont {Cerjan}, \citenamefont {Raman},\ and\ \citenamefont {Fan}}]{PhysRevLett.116.203902}%
  \BibitemOpen
  \bibfield  {author} {\bibinfo {author} {\bibfnamefont {A.}~\bibnamefont {Cerjan}}, \bibinfo {author} {\bibfnamefont {A.}~\bibnamefont {Raman}},\ and\ \bibinfo {author} {\bibfnamefont {S.}~\bibnamefont {Fan}},\ }\bibfield  {title} {\bibinfo {title} {Exceptional contours and band structure design in parity-time symmetric photonic crystals},\ }\href {https://doi.org/10.1103/PhysRevLett.116.203902} {\bibfield  {journal} {\bibinfo  {journal} {Phys. Rev. Lett.}\ }\textbf {\bibinfo {volume} {116}},\ \bibinfo {pages} {203902} (\bibinfo {year} {2016})}\BibitemShut {NoStop}%
\bibitem [{\citenamefont {Castro~Neto}\ \emph {et~al.}(2009)\citenamefont {Castro~Neto}, \citenamefont {Guinea}, \citenamefont {Peres}, \citenamefont {Novoselov},\ and\ \citenamefont {Geim}}]{RevModPhys.81.109}%
  \BibitemOpen
  \bibfield  {author} {\bibinfo {author} {\bibfnamefont {A.~H.}\ \bibnamefont {Castro~Neto}}, \bibinfo {author} {\bibfnamefont {F.}~\bibnamefont {Guinea}}, \bibinfo {author} {\bibfnamefont {N.~M.~R.}\ \bibnamefont {Peres}}, \bibinfo {author} {\bibfnamefont {K.~S.}\ \bibnamefont {Novoselov}},\ and\ \bibinfo {author} {\bibfnamefont {A.~K.}\ \bibnamefont {Geim}},\ }\bibfield  {title} {\bibinfo {title} {The electronic properties of graphene},\ }\href {https://doi.org/10.1103/RevModPhys.81.109} {\bibfield  {journal} {\bibinfo  {journal} {Rev. Mod. Phys.}\ }\textbf {\bibinfo {volume} {81}},\ \bibinfo {pages} {109} (\bibinfo {year} {2009})}\BibitemShut {NoStop}%
\bibitem [{\citenamefont {Tan}\ \emph {et~al.}(2007)\citenamefont {Tan}, \citenamefont {Zhang}, \citenamefont {Bolotin}, \citenamefont {Zhao}, \citenamefont {Adam}, \citenamefont {Hwang}, \citenamefont {Das~Sarma}, \citenamefont {Stormer},\ and\ \citenamefont {Kim}}]{PhysRevLett.99.246803}%
  \BibitemOpen
  \bibfield  {author} {\bibinfo {author} {\bibfnamefont {Y.-W.}\ \bibnamefont {Tan}}, \bibinfo {author} {\bibfnamefont {Y.}~\bibnamefont {Zhang}}, \bibinfo {author} {\bibfnamefont {K.}~\bibnamefont {Bolotin}}, \bibinfo {author} {\bibfnamefont {Y.}~\bibnamefont {Zhao}}, \bibinfo {author} {\bibfnamefont {S.}~\bibnamefont {Adam}}, \bibinfo {author} {\bibfnamefont {E.~H.}\ \bibnamefont {Hwang}}, \bibinfo {author} {\bibfnamefont {S.}~\bibnamefont {Das~Sarma}}, \bibinfo {author} {\bibfnamefont {H.~L.}\ \bibnamefont {Stormer}},\ and\ \bibinfo {author} {\bibfnamefont {P.}~\bibnamefont {Kim}},\ }\bibfield  {title} {\bibinfo {title} {Measurement of scattering rate and minimum conductivity in graphene},\ }\href {https://doi.org/10.1103/PhysRevLett.99.246803} {\bibfield  {journal} {\bibinfo  {journal} {Phys. Rev. Lett.}\ }\textbf {\bibinfo {volume} {99}},\ \bibinfo {pages} {246803} (\bibinfo {year} {2007})}\BibitemShut {NoStop}%
\bibitem [{\citenamefont {Tse}\ \emph {et~al.}(2008)\citenamefont {Tse}, \citenamefont {Hwang},\ and\ \citenamefont {Das~Sarma}}]{10.1063/1.2956669}%
  \BibitemOpen
  \bibfield  {author} {\bibinfo {author} {\bibfnamefont {W.-K.}\ \bibnamefont {Tse}}, \bibinfo {author} {\bibfnamefont {E.~H.}\ \bibnamefont {Hwang}},\ and\ \bibinfo {author} {\bibfnamefont {S.}~\bibnamefont {Das~Sarma}},\ }\bibfield  {title} {\bibinfo {title} {{Ballistic hot electron transport in graphene}},\ }\href {https://doi.org/10.1063/1.2956669} {\bibfield  {journal} {\bibinfo  {journal} {Applied Physics Letters}\ }\textbf {\bibinfo {volume} {93}},\ \bibinfo {pages} {023128} (\bibinfo {year} {2008})}\BibitemShut {NoStop}%
\bibitem [{\citenamefont {Breusing}\ \emph {et~al.}(2011)\citenamefont {Breusing}, \citenamefont {Kuehn}, \citenamefont {Winzer}, \citenamefont {Mali\ifmmode~\acute{c}\else \'{c}\fi{}}, \citenamefont {Milde}, \citenamefont {Severin}, \citenamefont {Rabe}, \citenamefont {Ropers}, \citenamefont {Knorr},\ and\ \citenamefont {Elsaesser}}]{PhysRevB.83.153410}%
  \BibitemOpen
  \bibfield  {author} {\bibinfo {author} {\bibfnamefont {M.}~\bibnamefont {Breusing}}, \bibinfo {author} {\bibfnamefont {S.}~\bibnamefont {Kuehn}}, \bibinfo {author} {\bibfnamefont {T.}~\bibnamefont {Winzer}}, \bibinfo {author} {\bibfnamefont {E.}~\bibnamefont {Mali\ifmmode~\acute{c}\else \'{c}\fi{}}}, \bibinfo {author} {\bibfnamefont {F.}~\bibnamefont {Milde}}, \bibinfo {author} {\bibfnamefont {N.}~\bibnamefont {Severin}}, \bibinfo {author} {\bibfnamefont {J.~P.}\ \bibnamefont {Rabe}}, \bibinfo {author} {\bibfnamefont {C.}~\bibnamefont {Ropers}}, \bibinfo {author} {\bibfnamefont {A.}~\bibnamefont {Knorr}},\ and\ \bibinfo {author} {\bibfnamefont {T.}~\bibnamefont {Elsaesser}},\ }\bibfield  {title} {\bibinfo {title} {Ultrafast nonequilibrium carrier dynamics in a single graphene layer},\ }\href {https://doi.org/10.1103/PhysRevB.83.153410} {\bibfield  {journal} {\bibinfo  {journal} {Phys. Rev. B}\ }\textbf {\bibinfo {volume} {83}},\ \bibinfo {pages} {153410} (\bibinfo {year} {2011})}\BibitemShut {NoStop}%
\bibitem [{\citenamefont {Wang}\ \emph {et~al.}(2010)\citenamefont {Wang}, \citenamefont {Strait}, \citenamefont {George}, \citenamefont {Shivaraman}, \citenamefont {Shields}, \citenamefont {Chandrashekhar}, \citenamefont {Hwang}, \citenamefont {Rana}, \citenamefont {Spencer}, \citenamefont {Ruiz-Vargas},\ and\ \citenamefont {Park}}]{10.1063/1.3291615}%
  \BibitemOpen
  \bibfield  {author} {\bibinfo {author} {\bibfnamefont {H.}~\bibnamefont {Wang}}, \bibinfo {author} {\bibfnamefont {J.~H.}\ \bibnamefont {Strait}}, \bibinfo {author} {\bibfnamefont {P.~A.}\ \bibnamefont {George}}, \bibinfo {author} {\bibfnamefont {S.}~\bibnamefont {Shivaraman}}, \bibinfo {author} {\bibfnamefont {V.~B.}\ \bibnamefont {Shields}}, \bibinfo {author} {\bibfnamefont {M.}~\bibnamefont {Chandrashekhar}}, \bibinfo {author} {\bibfnamefont {J.}~\bibnamefont {Hwang}}, \bibinfo {author} {\bibfnamefont {F.}~\bibnamefont {Rana}}, \bibinfo {author} {\bibfnamefont {M.~G.}\ \bibnamefont {Spencer}}, \bibinfo {author} {\bibfnamefont {C.~S.}\ \bibnamefont {Ruiz-Vargas}},\ and\ \bibinfo {author} {\bibfnamefont {J.}~\bibnamefont {Park}},\ }\bibfield  {title} {\bibinfo {title} {{Ultrafast relaxation dynamics of hot optical phonons in graphene}},\ }\href {https://doi.org/10.1063/1.3291615} {\bibfield  {journal} {\bibinfo  {journal} {Applied Physics Letters}\ }\textbf {\bibinfo {volume} {96}},\ \bibinfo {pages} {081917} (\bibinfo {year} {2010})}\BibitemShut {NoStop}%
\bibitem [{\citenamefont {Mihnev}\ \emph {et~al.}(2016)\citenamefont {Mihnev}, \citenamefont {Kadi}, \citenamefont {Divin}, \citenamefont {Winzer}, \citenamefont {Lee}, \citenamefont {Liu}, \citenamefont {Zhong}, \citenamefont {Berger}, \citenamefont {de~Heer}, \citenamefont {Malic}, \citenamefont {Knorr},\ and\ \citenamefont {Norris}}]{Mihnev2016}%
  \BibitemOpen
  \bibfield  {author} {\bibinfo {author} {\bibfnamefont {M.~T.}\ \bibnamefont {Mihnev}}, \bibinfo {author} {\bibfnamefont {F.}~\bibnamefont {Kadi}}, \bibinfo {author} {\bibfnamefont {C.~J.}\ \bibnamefont {Divin}}, \bibinfo {author} {\bibfnamefont {T.}~\bibnamefont {Winzer}}, \bibinfo {author} {\bibfnamefont {S.}~\bibnamefont {Lee}}, \bibinfo {author} {\bibfnamefont {C.-H.}\ \bibnamefont {Liu}}, \bibinfo {author} {\bibfnamefont {Z.}~\bibnamefont {Zhong}}, \bibinfo {author} {\bibfnamefont {C.}~\bibnamefont {Berger}}, \bibinfo {author} {\bibfnamefont {W.~A.}\ \bibnamefont {de~Heer}}, \bibinfo {author} {\bibfnamefont {E.}~\bibnamefont {Malic}}, \bibinfo {author} {\bibfnamefont {A.}~\bibnamefont {Knorr}},\ and\ \bibinfo {author} {\bibfnamefont {T.~B.}\ \bibnamefont {Norris}},\ }\bibfield  {title} {\bibinfo {title} {Microscopic origins of the terahertz carrier relaxation and cooling dynamics in graphene},\ }\href {https://doi.org/10.1038/ncomms11617} {\bibfield  {journal} {\bibinfo  {journal} {Nature Communications}\ }\textbf {\bibinfo {volume} {7}},\ \bibinfo {pages} {11617} (\bibinfo {year} {2016})}\BibitemShut {NoStop}%
\bibitem [{\citenamefont {Hanson}(2008)}]{10.1063/1.2891452}%
  \BibitemOpen
  \bibfield  {author} {\bibinfo {author} {\bibfnamefont {G.~W.}\ \bibnamefont {Hanson}},\ }\bibfield  {title} {\bibinfo {title} {{Dyadic Green's functions and guided surface waves for a surface conductivity model of graphene}},\ }\href {https://doi.org/10.1063/1.2891452} {\bibfield  {journal} {\bibinfo  {journal} {Journal of Applied Physics}\ }\textbf {\bibinfo {volume} {103}},\ \bibinfo {pages} {064302} (\bibinfo {year} {2008})}\BibitemShut {NoStop}%
\bibitem [{\citenamefont {Majumdar}\ \emph {et~al.}(2013)\citenamefont {Majumdar}, \citenamefont {Kim}, \citenamefont {Vuckovic},\ and\ \citenamefont {Wang}}]{Majumdar2013}%
  \BibitemOpen
  \bibfield  {author} {\bibinfo {author} {\bibfnamefont {A.}~\bibnamefont {Majumdar}}, \bibinfo {author} {\bibfnamefont {J.}~\bibnamefont {Kim}}, \bibinfo {author} {\bibfnamefont {J.}~\bibnamefont {Vuckovic}},\ and\ \bibinfo {author} {\bibfnamefont {F.}~\bibnamefont {Wang}},\ }\bibfield  {title} {\bibinfo {title} {Electrical control of silicon photonic crystal cavity by graphene},\ }\href {https://doi.org/10.1021/nl3039212} {\bibfield  {journal} {\bibinfo  {journal} {Nano Letters}\ }\textbf {\bibinfo {volume} {13}},\ \bibinfo {pages} {515} (\bibinfo {year} {2013})}\BibitemShut {NoStop}%
\bibitem [{\citenamefont {Satoshi}\ \emph {et~al.}(2023)\citenamefont {Satoshi}, \citenamefont {Otsuka}, \citenamefont {Moritake}, \citenamefont {Yoda}, \citenamefont {Uemura}, \citenamefont {Ono}, \citenamefont {Kuramochi},\ and\ \citenamefont {Notomi}}]{Satoshi:23}%
  \BibitemOpen
  \bibfield  {author} {\bibinfo {author} {\bibfnamefont {S.}~\bibnamefont {Satoshi}}, \bibinfo {author} {\bibfnamefont {S.}~\bibnamefont {Otsuka}}, \bibinfo {author} {\bibfnamefont {Y.}~\bibnamefont {Moritake}}, \bibinfo {author} {\bibfnamefont {T.}~\bibnamefont {Yoda}}, \bibinfo {author} {\bibfnamefont {T.}~\bibnamefont {Uemura}}, \bibinfo {author} {\bibfnamefont {M.}~\bibnamefont {Ono}}, \bibinfo {author} {\bibfnamefont {E.}~\bibnamefont {Kuramochi}},\ and\ \bibinfo {author} {\bibfnamefont {M.}~\bibnamefont {Notomi}},\ }\bibfield  {title} {\bibinfo {title} {Non-hermitian chirality and topological properties of graphene-loaded photonic crystals},\ }in\ \href {https://doi.org/10.1364/CLEO_FS.2023.FM4B.3} {\emph {\bibinfo {booktitle} {CLEO 2023}}}\ (\bibinfo  {publisher} {Optica Publishing Group},\ \bibinfo {year} {2023})\ p.\ \bibinfo {pages} {FM4B.3}\BibitemShut {NoStop}%
\bibitem [{\citenamefont {Shalaev}\ \emph {et~al.}(2019)\citenamefont {Shalaev}, \citenamefont {Walasik}, \citenamefont {Tsukernik}, \citenamefont {Xu},\ and\ \citenamefont {Litchinitser}}]{Shalaev2019}%
  \BibitemOpen
  \bibfield  {author} {\bibinfo {author} {\bibfnamefont {M.~I.}\ \bibnamefont {Shalaev}}, \bibinfo {author} {\bibfnamefont {W.}~\bibnamefont {Walasik}}, \bibinfo {author} {\bibfnamefont {A.}~\bibnamefont {Tsukernik}}, \bibinfo {author} {\bibfnamefont {Y.}~\bibnamefont {Xu}},\ and\ \bibinfo {author} {\bibfnamefont {N.~M.}\ \bibnamefont {Litchinitser}},\ }\bibfield  {title} {\bibinfo {title} {Robust topologically protected transport in photonic crystals at telecommunication wavelengths},\ }\href {https://doi.org/10.1038/s41565-018-0297-6} {\bibfield  {journal} {\bibinfo  {journal} {Nature Nanotechnology}\ }\textbf {\bibinfo {volume} {14}},\ \bibinfo {pages} {31} (\bibinfo {year} {2019})}\BibitemShut {NoStop}%
\bibitem [{\citenamefont {Sakoda}\ and\ \citenamefont {Sakoda}(2005)}]{sakoda2005optical}%
  \BibitemOpen
  \bibfield  {author} {\bibinfo {author} {\bibfnamefont {K.}~\bibnamefont {Sakoda}}\ and\ \bibinfo {author} {\bibfnamefont {K.}~\bibnamefont {Sakoda}},\ }\href@noop {} {\emph {\bibinfo {title} {Optical properties of photonic crystals}}},\ Vol.~\bibinfo {volume} {2}\ (\bibinfo  {publisher} {Springer},\ \bibinfo {year} {2005})\BibitemShut {NoStop}%
\end{thebibliography}%

\end{document}